\documentclass[a4paper,11pt,twoside]{article}
\usepackage[a4paper,top=3cm, bottom=3cm, left=3cm, right=3cm]{geometry}
\usepackage[american]{babel}
\usepackage[utf8]{inputenc}       
\usepackage[T1]{fontenc}
\usepackage{fourier}
\usepackage{bbm}
\usepackage{amssymb,amsmath}
\usepackage{amsthm}
\usepackage{latexsym}
\usepackage{authblk}
\usepackage[shortlabels]{enumitem}
\usepackage{etoolbox}
\usepackage{needspace}
\usepackage{graphicx,color,xcolor}
\usepackage{cite}
\usepackage[colorlinks=true,hyperindex=true]{hyperref}
\colorlet{darkblue}{blue!50!black}
\hypersetup{
	colorlinks,%
	citecolor=darkblue,%
	filecolor=red,%
	linkcolor=darkblue,%
	urlcolor=darkblue,%
	pdfnewwindow=true,%
	pdfstartview={FitH}
}
\newcommand{\tr}{\mathop{\mathrm{tr}}\nolimits}
\newcommand{\e}{\mathrm{e}}
\renewcommand{\d}{\mathrm{d}}

\renewcommand{\i}{\mathrm{i}}

\newcommand{\zz}{\mathbb{Z}}
\newcommand{\nn}{\mathbb{N}}
\newcommand{\rr}{\mathbb{R}}
\newcommand{\cc}{\mathbb{C}}

\newcommand{\cH}{\mathcal{H}}
\newcommand{\fA}{\mathfrak{A}}
\newcommand{\fAloc}{{\mathfrak{A}_\mathrm{loc}}}
\newcommand{\fh}{\mathfrak{h}}
\newcommand{\AAA}{\mathcal{A}}

\newcommand{\cS}{\mathcal{S}}
\newcommand{\cT}{\mathcal{T}}
\newcommand{\cF}{\mathcal{F}}
\newcommand{\cB}{\mathcal{B}}
\newcommand{\cBf}{\mathcal{B}_\mathrm{f}}
\newcommand{\cBr}{\mathcal{B}^r}
\newcommand{\cSI}{\mathcal{S}_\mathrm{I}}
\newcommand{\cSeq}{\mathcal{S}_\mathrm{eq}}
\newcommand{\cSwg}{\mathrm{WG}}
\newcommand{\cSewg}{\mathrm{WG}}
\newcommand{\cSlwg}{\mathrm{LWG}}
\newcommand{\cO}{\mathcal{O}}

\renewcommand{\Im}{\mathop{\mathrm{Im}}\nolimits}
\newcommand{\supp}{\mathop{\mathrm{supp}}\nolimits}
\newcommand{\diam}{\mathop{\mathrm{diam}}\nolimits}
\newcommand{\Dom}{\mathop{\mathrm{Dom}}\nolimits}
\newcommand{\CAR}{\mathop{\mathrm{CAR}}\nolimits}
\newcommand{\wstar}[1]{$\mathrm{weak}^\ast$\!-{#1}}
\def\one{\mathbbm{1}}
\def\be{\begin{equation}}
\def\ee{\end{equation}}

\theoremstyle{plain}
\newtheorem{thm}{Theorem}[section]
\newtheorem{cor}[thm]{Corollary}

\newtheorem{prop}[thm]{Proposition}

\theoremstyle{definition}
\newtheorem{defn}[thm]{Definition}

\AtBeginEnvironment{thm}{\Needspace{5\baselineskip}}
\AtBeginEnvironment{prop}{\Needspace{5\baselineskip}}
\AtBeginEnvironment{defn}{\Needspace{5\baselineskip}}

\usepackage{fancyhdr}
\usepackage{xargs}
\usepackage[textwidth=2.4cm]{todonotes}
\newcommandx{\ca}[2][1=]{\todo[inline,author={ca},
	linecolor=green,backgroundcolor=green!15,bordercolor=green,#1]{#2}}
\newcommandx{\canot}[2][1=]{\todo[
	linecolor=green,backgroundcolor=green!15,bordercolor=green,#1]{#2}}

\title{Approach to equilibrium in translation-invariant\\ quantum systems: some structural results}
\author[1]{Vojkan Jak\v si\'c}
\author[2]{Claude-Alain Pillet}
\author[3]{Clément Tauber}
\affil[1]{Department of Mathematics and Statistics, McGill University,
	805 Sherbrooke Street West, Montreal, QC, H3A 2K6, Canada}
\affil[2]{Aix Marseille Univ, Université de Toulon, CNRS, CPT, Marseille, France}
\affil[3]{Institut de Recherche Mathématique Avancée, UMR 7501 Université de Strasbourg et CNRS, 7 rue René-Descartes, 67000 Strasbourg, France}
\date{\today}
\begin{document}
\def\today{}
\maketitle
\bigskip
\centerline{\large \bf Dedicated to the memory of Krzysztof Gaw\k edzki}

\vskip1cm
\begin{small}
\noindent{\bf Abstract.}
We formulate the problem of approach to equilibrium in algebraic quantum statistical mechanics and study some of its structural aspects, focusing on the relation between the zeroth law of thermodynamics (approach to equilibrium) and the second law (increase of entropy). Our main result is that approach to equilibrium is necessarily accompanied by a \textit{strict} increase of the specific (mean) energy and entropy. In the course of our analysis, we introduce the concept of quantum weak Gibbs state which is of independent interest.
\end{small}

\section{Introduction}
\label{sec:intr}

Algebraic quantum statistical mechanics provides a general framework for
studying  dynamical aspects of infinitely extended quantum systems. Lattice
spins and fermions are examples of such systems which, in the translation
invariant setting, are also the main focus of this work. Adapting the algebraic 
framework, in this paper we initiate  a research program dealing with the
fundamental  problem of Boltzmann\footnote{For additional information and
references about this problem see for example~\cite{Lebowitz1993,Uhlenbeck1963}}
which is succinctly summarized in~\cite[Appendix I.1]{Simon1993} as follows:

\begin{quote}\textit{Approach to equilibrium: heuristic description.}
	
``As for the laws of thermodynamics, most textbooks primarily discuss the first
and the second law. Actually, there  is a fundamental experimental fact basic to
all thermodynamic descriptions that is usually implicitly assumed: It is
occasionally called the zeroth law\footnote{This terminology is unfortunate; 
see~\cite{Brown2001}} [\dots]. The zeroth law deals with the observed fact that a large
system seem to normally have `states' described by a few macroscopic parameters
like a temperature and density,  and that any system not in one of these states,
left alone,  rapidly approaches one of these states. When Boltzmann and Gibbs
tried to find a macroscopic basis for thermodynamics, they realized that the
approach to equilibrium was the most puzzling and deepest problem in such a
formalism.''
\end{quote}

Previous works on this topic in the algebraic formalism are scarce. Approach to
equilibrium was discussed in~\cite{Haag1973,Lanford1972,Sukhov1983} in the
context of quasi-free fermionic dynamics, and more generally for interacting
fermionic systems in~\cite{Erdos2004,Hugenholtz1983,Hugenholtz1987}. The quantum
Ising model was analyzed in~\cite{Radin1970}.

The goal of this work is to investigate some structural properties of the
quantum dynamical systems associated to interacting, infinitely extended lattice
spin systems and lattice Fermi gases. Besides setting the program, we examine
one important special case where the system, initially in thermal equilibrium
w.r.t.\;an interaction $\Psi$, evolves according to the dynamics generated by
some other interaction $\Phi$, focusing on the relation between the zeroth and
the second law  of thermodynamics. Our main result states that, in this setting,
approach to equilibrium is accompanied by a \textit{strict} increase of the
specific entropy and of the specific energy of the original interaction $\Psi$.
Let us emphasize the \textit{strict} increase aspect of our result. That the
entropy can't decrease is an easy consequence of the upper-semicontinuity of the
specific entropy and of the classical result of Lanford and
Robinson~\cite{Lanford1968} that this specific entropy remain constant along the
state trajectory. That the specific energy of $\Psi$ cannot decrease is a
consequence of the Gibbs variational principle and a general estimate of Hiai
and Petz~\cite[Lemma 2.3]{Hiai1993a}; see Theorem~\ref{thm-mil}. One compelling
aspect of our results is their generality.
 
Our analysis builds on the foundational works on statistical mechanics of
quantum spin systems that go back to 1970's (summarized in the
monographs~\cite{Ruelle1969,Bratteli1981,Israel1979,Simon1993}), and their
extensions to fermionic systems~\cite{Araki2003}. The two additional ingredients
are:
\begin{enumerate}[{\bf (\alph*)}]
\item The notion of weak Gibbs states.
\item The set of ideas that emerged during the last twenty years in the studies
of entropy production in the algebraic framework of non-equilibrium quantum
statistical mechanics; see~\cite{Jaksic2001a,Ruelle2000}.
\end{enumerate}
The notion of weak Gibbs states was implicit in the early works on thermodynamic
formalism of classical spin systems, but surprisingly it was formalized only
relatively late~\cite{Yuri2002}; see
also~\cite{Berghout2019,Fernandez1997,Kuelske2004,Jordan2011,Maes1999a,Pfister2020,Varandas2012,Enter2004}.
Similarly, the notion of quantum weak Gibbs state has been implicit in a number
of works in quantum statistical mechanics~\cite{Araki1974b,  Araki1976,Hiai1993a, Hiai2007,  Ogata2011}
but, to the best of our knowledge, it has not been formalized before. Besides
our work, the study of quantum weak Gibbs states is of independent interest, and
most of   foundational questions remain open.

Regarding~(b), and more generally in order to put the results of this paper and
the research program it initiates in a proper perspective, the next
section provides a short account of dynamical aspects of algebraic quantum
statistical mechanics, assuming that the reader has had a previous exposure to
the subject. The problem of approach to equilibrium in algebraic quantum
statistical mechanics is also formulated in this section.
 
\subsection{The algebraic approach to nonequilibrium quantum statistical mechanics}
\label{sec-setting}
 
Our starting point is a $C^\ast$-dynamical system $(\cO,\alpha)$, where $\cO$ is
the unital\footnote{The unit will be denoted by $\one$.} $C^\ast$-algebra of
observables of the quantum mechanical system under consideration. The Heisenberg
dynamics of this system is described by a strongly continuous one-parameter
subgroup $\alpha=\{\alpha^t\mid t\in\rr\}$ of the group $\mathrm{Aut}(\cO)$ of
$\ast$-automorphisms of $\cO$---we will often refer to such $\alpha$ as a
$C^\ast$-dynamics. States of the system are normalized positive linear
functionals on $\cO$. The set $\cS(\cO)$ of all states is a convex
\wstar{compact} subset of the dual Banach space $\cO^\ast$. In the specific case
of a system of spins or fermions on a lattice, we will denote the algebra of
observables by $\fA$; see Section~\ref{sec-lattice}. For some models the natural
starting point is rather a $W^\ast$-dynamical system; we will comment on such
models in Section~\ref{sec-discussion}.

The seminal work~\cite{Haag1967} introduced the KMS-condition as a
characterization of thermal equilibrium states of $(\cO,\alpha)$ at inverse
temperature $\beta>0$. A state $\omega\in\cS(\cO)$ is $(\alpha,\beta)$-KMS if,
for all $A, B\in\cO$, the function defined by
\[
\rr \ni t \mapsto F_{A, B}(t)=\omega(A\alpha^t(B)),
\] 
has an analytic continuation to the strip $0 <\Im z <\beta$, that is bounded and
continuous on its closure, and satisfies the KMS-boundary condition
\[
F_{A,B}(t+\i\beta)= \omega(\alpha^t(B)A).
\]
Any $(\alpha,\beta)$-KMS state is $\alpha$-invariant. A quantum dynamical system
$(\cO,\alpha,\omega)$, where $\omega$ is a $(\alpha,\beta)$-KMS state, describes
a physical system in thermal equilibrium at temperature $1/\beta$.

It is mathematically convenient to extend, in the obvious way, the definition of
KMS-states to all $\beta\in\rr$. Of particular interest is the value $\beta=-1$
which links the KMS-condition to the Tomita--Takesaki modular theory of operator
algebras.

The general theory of KMS states, developed in the early days of algebraic
quantum statistical mechanics, is summarized in~\cite[Sections~5.3
and~5.4]{Bratteli1981}. To set up the notation to be used below, let us mention
the following remarkable stability property of a $(\alpha,\beta)$-KMS state
$\omega\in\cS(\cO)$. Denoting by $\delta$ the generating derivation
$\alpha^t=\e^{t\delta}$, there is a norm-continuous map  $\cO\ni
V=V^\ast\mapsto\omega_V\in\cS(\cO)$, such that $\omega_0=\omega$ and $\omega_V$
is $(\alpha_V,\beta)$-KMS, where $\alpha_V$ is generated by
$\delta+\i[V,\,\cdot\,]$. We shall say that the $C^\ast$-dynamics $\alpha_V$ is
a local perturbation of $\alpha$.

Dynamical characterizations of KMS states are of particular relevance in the
context of the present work; see~\cite[Section~5.4.2]{Bratteli1981}. Indeed, the
problem of approach  to equilibrium can be viewed as an ultimate step in such
characterizations. More generally, the foundations of algebraic quantum
statistical mechanics are described in the classical monographs~\cite{Bratteli1987,
Bratteli1981,Haag1996,Israel1979,Simon1993,Ruelle1969,Thirring2002}; for a
more modern perspective see~\cite{Derezinski2003a,Jaksic2010b,Pillet2006}. 
From those foundations stem the two directions of research that were actively 
pursued over the last twenty-five years.
 
\begin{quote}\textit{1. Return to Equilibrium.} The formulation of this  property of
quantum dynamical systems  and the first basic results go back to the seminal
works of Robinson~\cite{Robinson1973, Robinson1976}; \cite{Botvich1983,Araki2003,
Jaksic1996a,Jaksic1997a,Bach2000,Jaksic2001b,Birke2002,Derezinski2003,Froehlich2003a,
Froehlich2004,DeRoeck2011} is an incomplete list of references of the follow-up works
on the subject. Among several equivalent formulations of return to equilibrium
we choose here the following one. The quantum dynamical system
$(\cO,\alpha,\omega)$, in thermal equilibrium at inverse temperature $\beta$, is
said to have the property of return to equilibrium if, for all $V=V^\ast\in\cO$
and all $A\in\cO$,
\[
\lim_{T\rightarrow \infty}\frac{1}{T}\int_0^T\omega_V\circ\alpha^t(A)\d t = \omega(A).
\]
In the setting of classical dynamical systems, the property of return to
equilibrium is equivalent to ergodicity, and its importance/relevance in quantum
statistical mechanics parallels the classical one.
\end{quote}
 
\begin{quote}\textit{2. Non-Equilibrium Steady States and Entropy Production.} This
topic  was implicit in some early works in algebraic quantum statistical
mechanics; see in particular~\cite{Pusz1978}. Its modern formulation goes back
to~\cite{Ruelle2000,Ruelle2001,Jaksic2001a} and builds on the notions of
non-equilibrium steady states and entropy production in classical statistical
mechanics introduced in early 1990's; see~\cite{Ruelle1999} for references and
additional information.  To describe this topic, consider a quantum dynamical
system $(\cO,\alpha,\omega)$ where  $\omega$ is $\alpha$-invariant but \textit{not}
$(\alpha, \beta)$-KMS  for any $\beta \in \rr$. The Non-Equilibrium Steady
States (NESS) of $(\cO,\alpha,\omega)$ associated to $V=V^\ast\in\cO$ are the
\wstar{limit} points of the net
\[
\left\{ \frac{1}{T}\int_0^T\omega \circ \alpha_V^t \d t\right\}_{T>0}
\]
as $T\uparrow \infty$. The set of NESS is non-empty and its elements are
$\alpha_V$-invariant. To define the entropy production of a NESS, we suppose
that there exists a reference $C^\ast$-dynamics $\varsigma_\omega$ such that
$\omega$ is a $(\varsigma_\omega, -1)$-KMS state. Let $\delta_\omega$ be the
generator of $\varsigma_\omega$, and suppose that $V$ is in the domain of
$\delta_\omega$. The associated entropy production observable is
$\sigma_V=\delta_\omega(V)$. The entropy production rate of a  NESS $\omega_+$
is the real number $\omega_+(\sigma_V)$. The pertinence of  this definition
stems from the entropy balance equation of~\cite{Pusz1978,Ruelle1978,Jaksic2001a}:
\begin{equation} S(\omega\circ \alpha_V^T|\omega)=\int_0^T \omega\circ \alpha_V^t(\sigma_V)\d t,
\label{ep-ness}
\end{equation}
where $S:\cS(\cO)\times\cS(\cO)\rightarrow[0,\infty]$ is Araki's relative
entropy functional.\footnote{Our sign and ordering conventions are such that,
for density matrices $\rho_1,\rho_2$, $S(\rho_1|\rho_2)=\tr(\rho_1(\log
\rho_1-\log \rho_2))$. We refer the reader to~\cite{Ohya1993} for an in depth
discussion of relative entropy.} In particular, the sign of $S$ and the
definition of NESS  immediately  give that $\omega_+(\sigma_V)\geq 0$. The 
works~\cite{Ho2000,Pillet2001,Jaksic2001a,Jaksic2002b,Jaksic2002a,Froehlich2003,
Jaksic2003,Matsui2003,Tasaki2003,Ogata2004,Aschbacher2006,Jaksic2006b,Jaksic2006c,
Jaksic2006a,Pillet2006,Aschbacher2007,Jaksic2007a,Merkli2007,Merkli2007a,
DeRoeck2009,Jaksic2010b,Jaksic2012} build on these starting points and develop
the structural theory of NESS and entropy production, including studies of several
classes of concrete physically relevant models.
\end{quote}

In this paper we initiate a third direction of research that also originates in
the early foundations of algebraic quantum statistical mechanics, and closes the
circle of ideas and techniques introduced in the study of the above two
directions.
 
\begin{quote}\textit{3. The problem of approach to equilibrium in quantum statistical
mechanics.} Consider a quantum dynamical system $(\fA, \alpha_\Phi, \omega)$,
where $\fA$ is the $C^\ast$-algebra of a system of lattice spins or fermions,
$\alpha_\Phi$ is the $C^\ast$-dynamics generated by  a sufficiently regular
translation invariant interaction $\Phi$, and $\omega$ is a translation
invariant state. The Equilibrium Steady States (ESS) of $(\fA,\alpha_\Phi,\omega)$
are the \wstar{limit} points of the net
\[
\left\{ \frac{1}{T}\int_0^T\omega\circ\alpha_\Phi^t\d t\right\}_{T>0}
\]
as $T\uparrow\infty$. The set of ESS is non-empty and its elements are
$\alpha_\Phi$-invariant. The aim is to develop the structural theory of ESS and
eventually determine the conditions under which ESS are KMS-states for
$\alpha_\Phi$.
\end{quote}
 
In this work we examine the structural aspects of the problem of approach to
equilibrium in the special case where $\omega$ is a KMS state for some
interaction $\Psi$. We show that, except in trivial cases, the process of
approach to equilibrium is accompanied by a strict increase of the specific
energy and entropy. To establish that, we introduce the concept of weak
Gibbsianity in quantum statistical mechanics and adapt the idea of entropy
balance to the approach to equilibrium setting.

\medskip
The remaining parts of this article are organized as follows. In
Section~\ref{sec-lattice}, we give a telegraphic review of quantum spin and
fermion systems. Weak Gibbs states and the related entropy balance equation are
discussed in Section~\ref{sec-weak-gibbs}. Our main results are stated in
Section~\ref{sec-main-results} and discussed in Section~\ref{sec-discussion}.

The proofs are given in Section~\ref{sec-proofs}.

In the follow-up work~\cite{Jaksic2023a}, we use the same general  ingredients
(a) and (b) above to examine the status of the adiabatic theorem in the
translation invariant setting of  infinitely extended lattice quantum spin and
fermionic systems.

\paragraph*{Acknowledgments} This work was supported by the French \textit{Agence
	Nationale de la Recherche}, grant NONSTOPS (ANR-17-CE40-0006-01,
ANR-17-CE40-0006-02, ANR-17-CE40-0006-03) and CY Initiative of Excellence, {\sl
	Investissements d'Avenir} program (grant ANR-16-IDEX-0008). It was partly
developed during VJ's stay at the CY Advanced Studies, whose support is
gratefully acknowledged. VJ also acknowledges the support of NSERC. The authors
wish to thank Laurent Bruneau and Aernout van Enter for useful discussions, and
Luc Rey-Bellet for pointing them to the reference~\cite{Moriya2020}.


\section{Setting and main results}
\subsection{Quantum spin and fermion systems}
\label{sec-lattice}
 
\paragraph{The spin $\boldsymbol{C^\ast}$-algebra.}
Let $\cH$ be the finite dimensional Hilbert space of a single spin   and
consider the lattice $\zz^d$ equipped with the norm $|x|=\sum_{j=1}^d|x_j|$. Let
$\cF$ be  the collection of all finite subsets of $\zz^d$.  We denote by
$\diam(X)=\max\{|x-y|\mid x,y\in X\}$ the diameter of $X\in\cF$, and write
$\cH_X=\bigotimes_{x\in X}\cH_x$, where $\cH_x=\cH$, and
$\fA_X=\cB(\cH_X)$.\footnote{$\cB(\cH)$ denotes the $C^\ast$-algebra of all
linear operators on the Hilbert space $\cH$.} If $Y\subset X$, we identify in
a natural way $\fA_{Y}$ with a subalgebra of $\fA_{X}$.  For $X\subset \zz^d$, 
we denote by ${\fA}_X$ the inductive limit $C^\ast$-algebra over the family
$\{\fA_Y\}_{X\supset Y\in\cF}$.  The kinematical algebra of the spin system is
${\fA}_{\zz^d}$ and in the sequel we will omit the subscript $\zz^d$. The
algebra $\fA$ is simple, which implies in particular that any KMS state on $\fA$
is faithful; see~\cite[Lemma III.3.3]{Israel1979}.

For any $X\subset \zz^d$, $X^c = \zz^d \setminus X$, we have the
identification $\fA=\fA_X\otimes \fA_{X^c}$, and $\tr_X:\fA\mapsto \fA_{X^c}$ 
denotes the usual normalized partial trace. An element $A\in {\fA}$ is called
\textit{local} if $A\in {\fA}_X$ for some $X\in\cF$. The minimal $X$ for which
this hold is called the support of $A$ and is denoted by $\supp(A)$. The set
$\fAloc$ of local elements is a dense $\ast$-subalgebra of $\fA$.

In the following $\Lambda$ will always denote a cube in $\zz^d$ centered at the
origin and $\Lambda\uparrow \zz^d$ denotes the limit over an increasing family
of such cubes.
 
\paragraph{Translation invariant states.} The spin algebra $\fA$ is uniformly
asymptotically Abelian w.r.t.\;the natural group action $\varphi:\zz^d\ni
x\mapsto\varphi^x\in\mathrm{Aut}(\fA)$, \textit{i.e.,}
\[
\lim_{|x|\rightarrow \infty}\| [\varphi^x(A),B]\|=0
\]
for all $A, B\in \fA$. We denote by $\cSI(\fA)$ the set of all translation
invariant, \textit{i.e.,} $\varphi$-invariant states on $\fA$. $\cSI(\fA)$ is a
\wstar{compact} convex subset of $\cS(\fA)$ and a Choquet simplex.

\paragraph{Specific entropies.} The restriction $\nu_\Lambda$ of a state
$\nu\in\cSI(\fA)$ to $\fA_\Lambda$ is represented by a density matrix in
$\fA_\Lambda$ which we denote by the same letter: For any $A\in\fA_\Lambda$,
$$
\nu(A)=\nu_\Lambda(A)=\tr(\nu_\Lambda A).
$$
The von~Neumann entropy of $\nu_\Lambda$ is defined by
$S(\nu_\Lambda)=-\tr(\nu_\Lambda \log \nu_\Lambda)$ and is non-negative. The
\textit{specific entropy} of $\nu$ is given by the limit
\be
s(\nu)=\lim_{\Lambda \uparrow \zz^d}\frac{S(\nu_\Lambda)}{|\Lambda|},
\label{eq:sDef}
\ee
which exists and defines an affine, \wstar{upper}-semicontinuous map
$\cSI(\fA)\ni\nu\mapsto s(\nu)$, taking its values in $[0,\log\dim\cH]$;
see~\cite[Proposition~6.2.38]{Bratteli1981}.

The \textit{specific relative entropy} of $\nu$ w.r.t.\;$\omega\in\cSI(\fA)$ is
defined by
\[
s(\nu|\omega)=\lim_{\Lambda\uparrow\zz^d}\frac{S(\nu_\Lambda|\omega_\Lambda)}{|\Lambda|},
\]
whenever the limit exists, in which case we have $s(\nu|\omega)\geq0$. 
 
\paragraph{Interactions.} An \textit{interaction} is a family
$\Phi=\{\Phi(X)\}_{X\in\cF}$, where $\Phi(X)\in\fA_X$ is self-adjoint. The
interaction $\Phi$ is translation invariant if
\[
\varphi^x(\Phi(X))=\Phi(X+x)
\]
holds for all $X\in\cF$ and $x\in\zz^d$. In what follows we consider only
translation invariant interactions. The \textit{local Hamiltonians} 
associated to $\Phi$ are defined by 
\[
H_\Lambda(\Phi) =\sum_{X\subset \Lambda}\Phi(X).
\] 
For $r>0$ we denote by $\cBr$ the set of all translation invariant interactions
such that
\be
\|\Phi\|_r=\sum_{X\ni 0}\e^{r (|X|-1)}\|\Phi(X)\|<\infty,
\label{eq:BrNorm}
\ee
where $|X|$ denotes the cardinality of $X\in\cF$. The pair $(\cBr,\|\cdot\|_r)$
is a Banach space. An interaction $\Phi$ is called \textit{finite range}
whenever, for some $m\in \nn$, ${\rm diam}(X)>m$ implies $\Phi(X)=0$. The set 
$\cBf$ of all finite range interactions is a dense subspace of $\cBr$.
 
In what follows we restrict ourselves to interactions in $\cBr$ although many of
the stated results hold in more general settings.
 
\paragraph{The Gibbs variational principle.}\cite[Proposition~6.2.39 and Theorem~6.2.40]{Bratteli1981} For any interaction $\Phi\in\cBr$ the limit
\begin{equation}\label{eq:def_pressure}
P(\Phi)=\lim_{\Lambda\uparrow \zz^d}\frac{1}{|\Lambda|}\log\tr(\e^{-H_\Lambda(\Phi)})
\end{equation}
exists and is finite. The function $P$ defined in this way is called the \textit{pressure}. Setting 
\[
E_\Phi=\sum_{X\ni0}\frac{\Phi(X)}{|X|}\in \fA,
\]
one has
\be
\lim_{\Lambda \uparrow \zz^d}
\frac{1}{|\Lambda|}\left\|H_\Lambda(\Phi)-\sum_{x\in\Lambda}\varphi^x(E_\Phi)\right\|=0,
\label{eq:aschball}
\ee
and in particular
\be
\lim_{\Lambda \uparrow \zz^d}
\frac{1}{|\Lambda|}\omega\left(H_\Lambda(\Phi)\right)=\omega(E_\Phi),
\label{eq:seForm}
\ee
for any $\omega\in\cSI(\fA)$. Thus, $E_\Phi$ can be considered as the
\textit{specific energy} observable of the interaction $\Phi$.

\begin{thm}[Gibbs variational principle]\label{thm:Gibbs_variational}
For any  $\beta\in\rr$ and $\Phi\in\cBr$,
\[
P(\beta\Phi)=\sup_{\nu\in\cSI(\fA)}\left(s(\nu)-\beta\nu(E_\Phi)\right).
\]
Moreover,
\[
\cSeq(\beta \Phi)=\left\{\nu\in\cSI(\fA)\mid P(\beta\Phi)=s(\nu)-\beta\nu(E_\Phi)\right\}
\]
is a non-empty convex compact subset of $\cSI(\fA)$. 
\end{thm}

The elements of $\cSeq(\beta \Phi)$ are called \textit{equilibrium states} 
for the interaction $\beta\Phi$.

\paragraph{Dynamics.}\cite[Theorem~7.6.2]{Ruelle1967} For an interaction 
$\Phi\in\cBr$ and a finite cube $\Lambda$, set
\[
\alpha_{\Phi,\Lambda}^t(A)=\e^{\i t H_\Lambda(\Phi)}A\e^{-\i t H_\Lambda(\Phi)}
\]
with $t\in\rr$ and $A\in\fAloc$. Then, the limit 
\[
\alpha_\Phi^t(A)=\lim_{\Lambda \uparrow \zz^d}\alpha_{\Phi,\Lambda}^t(A)
\]
exists and is uniform for $t$ in compact sets. Moreover, the family of maps
$\alpha_{\Phi}=\{\alpha_\Phi^t\mid t\in \rr\}$, originally defined on $\fAloc$,
extends uniquely to a $C^\ast$-dynamics on $\fA$. The group actions
$\alpha_\Phi$ and $\varphi$ commute.

We recall  the following basic result~\cite[Theorem~6.2.42]{Bratteli1981}.
\begin{thm} Suppose that $\Phi\in\cBr$. Then, for any $\beta\in\rr$, $\cSeq(\beta\Phi)$ 
coincides with the set of all translation invariant $(\alpha_\Phi,\beta)$-KMS states.
\end{thm}
 
The proofs of the above results give additional information. In particular, the
following result follows from the proofs of~\cite[Lemma~7.6.1 and
Theorem~7.6.2]{Ruelle1969} or~\cite[Theorem~6.2.4]{Bratteli1981}:

\begin{thm}\label{thm:y-k-1} Suppose that  $\Phi\in\cBr$. Then, for all $A\in\fAloc$, the map 
\begin{equation*}
\rr \ni t \mapsto \alpha_\Phi^t(A)\in \fA
\end{equation*}
has an analytic extension to the strip 
\[
|\Im z| <\frac{r}{2 \|\Phi\|_{r}},
\]
and  for $z$ in this strip 
\begin{equation*} 
\|\alpha_\Phi^z(A)\|\leq  \|A\|\e^{r |\supp(A)|} \frac{r }{r-2\|\Phi\|_r|\Im z| }.
\end{equation*}
\end{thm}
\medskip\noindent{\bf Remark.} According to~\cite[Theorem~6.2.4]{Bratteli1981},
the conclusions of Theorem~\ref{thm:y-k-1} still hold for inter\-actions $\Phi$
which lack translation invariance, provided the norm~\eqref{eq:BrNorm} is
replaced by
$$
\|\Phi\|_r=\sum_{N\ge1}\sup_{x\in\zz^d}\sum_{X\ni x\atop|X|=N}\|\Phi(X)\|\e^{r(|X|-1)}<\infty.
$$

\paragraph{Setting $\boldsymbol{\beta=1}$.} To reduce the number of parameters,
it is convenient to absorb $\beta$ into the inter\-action $\Phi$. With this
convention small inter\-actions correspond to high temperatures. We  will
comment further on this point at the end of Section~\ref{sec-main-results}.

In the remaining part of the paper we set  $\beta=1$. $(\alpha_\Phi,1)$-KMS
state will be abbreviated as $\alpha_\Phi$-KMS state.

\paragraph{Uniqueness of equilibrium state.}
\begin{thm}\label{thm:kms-uniqueness} Suppose that $\Phi\in\cBr$ and that either:
\begin{enumerate}[label=(\arabic*)]
\item $d=1$ and $\displaystyle\sum_{X\ni0}{\rm diam}(X)\|\Phi(X)\|<\infty$.
\item $d\geq 1$,  $r>\log \dim \cH$, and $\displaystyle\|\Phi\|_{r}< \frac{1}{2 \dim \cH}$.
\end{enumerate}
Then $\cSeq(\Phi)$ is a singleton. 
\end{thm}
Part~(1) is the classical result of Araki; see~\cite[Theorem~IV.6.1]{Simon1993}.
Regarding Part~(2), among the many results in the literature that establish the
uniqueness of the equilibrium state in the high temperature regime, we have
quoted here the one obtained relatively recently in~\cite{Froehlich2015}.

\paragraph{The Gibbs condition.}\cite[Proposition~6.2.17]{Bratteli1981}  We will
only introduce the part of the Gibbs condition that will be needed in the
sequel. For an interaction $\Phi\in\cBr$, the \textit{surface energies}
\begin{equation}
W_\Lambda(\Phi)=\sum_{X \cap \Lambda\not=\emptyset\atop X\cap \Lambda^c\not=\emptyset}\Phi(X)
\label{eq:surface-en}
\end{equation}
are well defined and satisfy 
\be
\lim_{\Lambda \uparrow \zz^d} \frac{\|W_\Lambda(\Phi)\|}{|\Lambda|}=0.
\label{eq:NoSurf}
\ee
 
\begin{thm}\label{thm:spin-Gibbs-con}  Let $\omega$ be a $\alpha_\Phi$-KMS
state, and  $\omega_{-W}$  the perturbed KMS-state with $W=W_\Lambda(\Phi)$.
Then  for any $\Lambda$ and any $A\in \fA_\Lambda$ one has
$$
\omega_{-W}(A) = \frac{\tr(\e^{-H_\Lambda(\Phi)}A)}{\tr(\e^{-H_\Lambda(\Phi)})}.
$$
\end{thm}

\paragraph{Physical equivalence.} The concept of physical equivalence was first
introduced in~\cite{Roos1974} as a physical interpretation of a technical
assumption from~\cite{Griffiths1971}, and was further developed
in~\cite{Israel1979}. Our definition below differs from the original one and is
based on~\cite[Section 4.7]{Ruelle1978}; for additional information
see~\cite[Section 2.4.6]{Van_Enter1993}.
\begin{defn}\label{def:phys_eq} Two interactions $\Phi, \Psi\in\cBr$ are called 
\textit{physically equivalent}, denoted $\Phi\sim \Psi$, if 
\[
\lim_{\Lambda \uparrow \zz^d}\frac{1}{|\Lambda|}
\sum_{x\in \Lambda}\varphi^x(E_{\Phi-\Psi})=\tr(E_{\Phi-\Psi})\one.
\]
\end{defn}
Notice that $E_{\Phi-\Psi} = E_\Phi - E_\Psi$. Physical equivalence is obviously 
an equivalence relation on $\cBr$. Moreover, if $\Phi_1\sim \Psi_1$ and 
$\Phi_2\sim \Psi_2$, then $a\Phi_1 + b\Phi_2\sim a\Psi_1 +  b\Psi_2$ for any 
$a, b\in \rr$.
\begin{thm}\label{thm:spin-phys-eq} Let $\Phi,\Psi\in\cBr$. The following 
statements are equivalent:
\begin{enumerate}[label={\rm (\alph*)}]
\item $\Phi\sim\Psi$. 
\item $\displaystyle\lim_{\Lambda\uparrow\zz^d}\frac{1}{|\Lambda|}\big(H_\Lambda(\Phi)-H_\Lambda(\Psi)\big)=\tr(E_{\Phi-\Psi})\one$.
\item The pressure $P$ is linear on the line segment connecting $\Phi$ and $\Psi$.
\item $\cSeq(\Phi)\cap\cSeq(\Psi)\not=\emptyset$.
\item $\cSeq(\Phi)=\cSeq(\Psi)$.
\item $\alpha_\Phi=\alpha_\Psi$.
\item $\displaystyle\sum_{X\in\cF}[\Phi(X)-\Psi(X),A]=0$ for all $A\in\fAloc$.
\end{enumerate}
\end{thm}
 
The equivalence between~(c), (d), (e), and~(f) is proved in~\cite[Theorem~III.4.2]{Israel1979}.
We prove the remaining parts in Section~\ref{sec-phys-eq} below.

\paragraph{The case of fermions.} We denote by $\AAA=\CAR(\fh)$ the CAR algebra
over the Hilbert space $\fh$. As usual, $a^\ast(f)/a(f)$ are the
creation/annihilation operators associated to $f\in\fh$, and $a^\#$ stands for
either $a$ or $a^\ast$. Given a unitary operator $u$ on $\fh$, the map
$a^\#(f)\mapsto a^\#(uf)$ uniquely extends to a $\ast$-automorphism $\alpha_u$
of $\AAA$, the \textit{Bogoliubov automorphism} generated by $u$. A strongly
continuous one-parameter group of unitaries $t\mapsto u^t$ on $\fh$ generates a
strongly continuous one-parameter subgroup of $\mathrm{Aut}(\AAA)$. The gauge
group of $\AAA$ is the group of Bogoliubov automorphisms
$\theta\mapsto\vartheta^\theta$ generated by $\e^{\i\theta\one}$. Physical
observables are gauge invariant, \textit{i.e.,} elements of
$$
\fA(\fh)=\{A\in\AAA\mid
\vartheta^\theta(A)=A\text{ for all }\theta\in\rr\},
$$
which coincides with the $C^\ast$-algebra generated by $\{a^\ast(f)a(g)\mid
f,g\in\fh\}\cup\{\one\}$. We will only consider the case $\fh = \ell^2(\mathbb
Z^d)$ and write $\fA$ for $\fA(\ell^2(\mathbb Z^d))$. For $X \subset \mathbb
Z^d$ we set $\fA_X=\fA(\ell^2(X))$ and identify $\fA_X$ with the
$C^*$-subalgebra of $\fA$ generated by
$$
\{a^\ast(f)a(g)\mid f,g\in\ell^2(\mathbb Z^d),\supp(f)\cup\supp(g)\subset X\}\cup\{\one\}.
$$
Note that if $X \cap Y=\emptyset$, then $\fA_{X}$ and $\fA_{Y}$ commute. The
natural unitary group action of $\zz^d$ on $\ell^2(\mathbb Z^d)$ gives rise to a
group of Bogoliubov automorphisms $\varphi^x$ such that
$\varphi^x(\fA_X)=\fA_{X+x}$.

For $X\in\cF$, $\AAA_X=\CAR(\ell^2(X))$ is isomorphic to the full matrix-algebra
$\cB(\cc^{2^{|X|}})$, so that $\AAA=\CAR(\ell^2(\mathbb Z^d))$ has a unique
tracial state $\tr$ as the extension of the unique tracial states of $\AAA_X$,
$X\in\cF$. $\AAA$ is isomorphic to the spin algebra,  with the main difference
that the identification  of $\AAA$ with $\AAA_X \otimes \AAA_{X^c}$ does not
hold anymore. Nevertheless, there exists an analogue of partial trace for
fermions called conditional expectation in~\cite[Theorem 4.7]{Araki2003}: for
$X\subset\zz^d$, there exists a projection $\cT_X : \AAA \to \AAA_X$ such that
$\tr(AB) = \tr(\cT_X(A)B)$ for any $B \in \AAA_X$.

All the above definitions and results stated for the spin algebra extend to the
gauge-invariant sector $\fA$ of the fermionic algebra $\AAA$. Most, but not all,
of these extensions are straightforward: for the Gibbs variational principle and
the Gibbs condition see~\cite[Theorems 11.4 and 7.5]{Araki2003}. In the
fermionic case there is also a special one-body interaction $N\in\cBf$, related
to gauge invariance, and given by
$$
N(X)=\begin{cases}
a_x^\ast a_x&\text{ if }X = \{x\};\\
0&\text{ otherwise}.
\end{cases} 
$$
We denote $N_\Lambda=\sum_{X\subset\Lambda}N(X)$ the local Hamiltonians
associated to $N$.\footnote{The interaction $N$ generates the gauge group,
$\vartheta^\theta=\alpha_N^\theta$.}  For $\omega\in S_I(\fA)$,
$$
\lim_{\Lambda\uparrow\mathbb Z^d} \dfrac{1}{|\Lambda|}\omega(N_\Lambda)=\omega(E_N),
$$
where $E_N=a^\ast_0a_0$. The perturbed interactions $\Phi-\mu N$ introduces
a chemical potential $\mu$.

To summarize, all results of this paper apply to both spin and fermions, and so
in the following $\fA$ refers equivalently to the spin or the gauge-invariant
sector of the CAR algebra, unless stated explicitly.

We finish this section with: 
\paragraph{Conservation laws.}

\begin{thm}\label{ver-tr}
For any $\omega\in\cSI(\fA)$ and any $\Phi\in\cBr$ the following hold for all $t\in \rr$:
\begin{enumerate}[label=(\arabic*)]
\itemsep0em	
\item  $s(\omega)=s(\omega \circ \alpha_\Phi^t)$.	
\item $\omega(E_\Phi)=\omega\circ \alpha_\Phi^t(E_\Phi)$.
\item In the fermionic case, $\omega(E_N)=\omega\circ \alpha_\Phi^t(E_N)$.
\end{enumerate}
\end{thm}
Part~(1) is due to~\cite{Lanford1968}. The proofs of the remaining statements
are given in Section~\ref{sec-proof-invar}.

\subsection{Weak Gibbs states and regularity}
\label{sec-weak-gibbs}
We set
\[
\omega_{\Phi,\Lambda}^\mathrm{c}=\e^{-H_\Lambda(\Phi)-P_\Lambda(\Phi)},\qquad 
P_\Lambda(\Phi)=\log\tr(\e^{-H_\Lambda(\Phi)}).
\]
\begin{defn}\label{def:wg}
A state $\omega\in\cSI(\fA)$ is called \textit{weak Gibbs} for the
interaction $\Phi\in\cBr$ if there exist constants $c_\Lambda >0$ satisfying
\begin{equation}
\lim_{\Lambda \uparrow \zz^d}\frac{1}{|\Lambda|}c_\Lambda =0,
\label{mil-er}
\end{equation}
and such that, for all $\Lambda$,
\be
\e^{-c_\Lambda}\omega_{\Phi,\Lambda}^\mathrm{c}\leq\omega_\Lambda\leq \e^{c_\Lambda} \omega_{\Phi,\Lambda}^\mathrm{c}.
\label{eq:ewgDef}
\ee
$\omega$ is called \textit{log weak Gibbs} for $\Phi$ if there exist
constants $c_\Lambda >0$ satisfying~\eqref{mil-er} and such that, for all
$\Lambda$,
\be
- c_\Lambda+\log \omega_{\Phi,\Lambda}^\mathrm{c}\leq\log\omega_\Lambda
\leq c_\Lambda+\log\omega_{\Phi,\Lambda}^\mathrm{c}.
\label{eq:lwgDef}
\ee
We denote by $\cSwg(\Phi)$ the set of all  weak Gibbs states for
$\Phi$, and by $\cSlwg(\Phi)$ the set of all log weak Gibbs states.
\end{defn}

The next proposition, whose proof is given in
Section~\ref{sec:proofweakGibbs_are_eq}, collects a few elementary properties of
the sets $\cSewg(\Phi)$ and $\cSlwg(\Phi)$.
\begin{prop}\label{prop:weakGibbs_are_eq}
For $\Phi,\Psi\in\cBr$, the following hold:
\begin{enumerate}[label=(\arabic*)]
\item $\omega\in\cSlwg(\Phi)$ $\Longleftrightarrow$ $\displaystyle\limsup_{\Lambda\uparrow\zz^d}|\Lambda|^{-1}\|\log\omega_\Lambda-\log\omega_{\Phi,\Lambda}^c\|=0$.
\item $\cSewg(\Phi)\subset\cSlwg(\Phi)\subset\cSeq(\Phi)$.
\item Either $\cSlwg(\Phi)=\cSlwg(\Psi)$ or $\cSlwg(\Phi)\cap\cSlwg(\Psi)=\emptyset$.
\item $\Phi\sim\Psi$ $\Longrightarrow$ $\cSlwg(\Phi)=\cSlwg(\Psi)$. Moreover, 
if\/ $\cSlwg(\Phi)=\cSlwg(\Psi)\neq\emptyset$, then $\Phi\sim\Psi$.
\end{enumerate}
\end{prop}

Regarding possible equalities in Part~(2), one has the following results.
\begin{thm}\label{thm:weakGibbs}
Suppose that either:
\begin{enumerate}[label=(\alph*)]
\itemsep0em
\item $d=1$ and $\Phi\in\cBf$;
\item $d\geq 1$, $\Phi\in\cBr$,  and $\|\Phi\|_r<r$.
\end{enumerate}
Then $\cSewg(\Phi)= \cSlwg(\Phi)=\cSeq(\Phi)$. 
\end{thm}
Section~\ref{secAraki} is devoted to the proof which is based on general results
from~\cite{Araki1969} and \cite[Section 3.1 and proof of Theorem 2]{Araki1974b},
see also~\cite{Araki1976} and \cite{Lenci2005}.

The following characterization of $\cSewg(\Phi)$, based on the Gibbs condition,
is proved in Section~\ref{sec-proof-wg}.
\begin{prop}\label{prop:w-s-1}
Suppose that $\Phi\in\cBr$ and let $\omega\in\cSeq (\Phi)$. Set 
\[
\underbar{d}_\Lambda=\inf_{A\in\fA_\Lambda\atop A>0}\frac{\omega(A)}{\omega_{-W_\Lambda(\Phi)}(A)},
\qquad  
\overline{d}_\Lambda=\sup_{A\in \fA_\Lambda\atop A>0}\frac{\omega(A)}{\omega_{-W_\Lambda(\Phi)}(A)}.
\]
Then $\omega\in\cSewg(\Phi)$ iff
\begin{equation}
\lim_{\Lambda\uparrow\zz^d}\frac{1}{|\Lambda|}\log\underbar{d}_\Lambda
=\lim_{\Lambda\uparrow\zz^d}\frac{1}{|\Lambda|}\log\overline{d}_\Lambda=0.
\label{ajde}
\end{equation}	
\end{prop}

We finish this brief discussion of quantum weak Gibbs states with a
characterization of $\cSlwg(\Phi)$.
\begin{prop}\label{mil-la}
Suppose that $\Phi\in\cBr$ and let $\omega\in\cSeq (\Phi)$. 
\begin{enumerate}[label=(\arabic*)]
\item For any $\nu \in \cS(\fA)$ and all $\Lambda\in\cF$, 
\[
S(\nu_\Lambda|\omega_\Lambda)- S(\nu_\Lambda|\omega_{\Phi,\Lambda}^\mathrm{c})\leq 2\|W_\Lambda(\Phi)\|.
\]
\item For $\Lambda\in\cF$, set 
\[
D_\Lambda=\inf_{\nu \in \cS(\fA) }\left(S(\nu_\Lambda|\omega_\Lambda)- S(\nu_\Lambda|\omega_{\Phi,\Lambda}^\mathrm{c})\right).
\]
Then $\omega\in\cSlwg(\Phi)$ iff
\begin{equation}
\lim_{\Lambda\uparrow\zz^d}\frac{D_\Lambda}{|\Lambda|}=0.
\label{ajde-mil}
\end{equation}	
\end{enumerate}
\end{prop}
Part~(1) is easily seen to be equivalent to the statement that for all $\Lambda\in\cF$, 
\[
\log \omega_{\Phi,\Lambda}^\mathrm{c}- \log \omega_\Lambda \leq 2\|W_\Lambda(\Phi)\|,
\]
which is proven in~\cite[Lemma 2.3]{Hiai1993a}. With $D_\Lambda$ as defined in Part~(2), one has
$$
D_\Lambda\leq\log \omega_{\Phi,\Lambda}^\mathrm{c}- \log \omega_\Lambda,
$$
so that Part~(2) is an immediate consequence of Part~(1), Relation~\eqref{eq:NoSurf} and the definition of log weak Gibbs states.

\medskip
For our purposes, the importance of weak Gibbs  states stems from the following
result that is an immediate consequence of Definition~\ref{def:wg}.
 
\begin{prop} \label{prop:weak-gibbs-regular} 
Let $\Phi\in\cBr$ and $\omega\in\cSlwg(\Phi)$. Then for any $\nu\in\cSI(\fA)$, 
\begin{equation}
s(\nu|\omega)= -s(\nu)+ \nu(E_\Phi)+ P(\Phi).
\label{eq:regularity_condition}
\end{equation}
In particular, $s(\nu|\omega)=0$ iff $\nu\in\cSeq(\Phi)$. 
\end{prop}

We will comment further on weak Gibbsianity and regularity in Section~\ref{sec-discussion}.

Since the validity of~\eqref{eq:regularity_condition} plays a central role in
our arguments, we single it out in:
 
\begin{defn}\label{def:regular}
A pair $(\omega,\Phi)\in\cSI(\fA)\times\cBr$ is called regular whenever
Relation~\eqref{eq:regularity_condition} holds for any $\nu\in\cSI(\fA)$.
\end{defn}

\noindent{\bf Remarks.} 1. Obviously, if the pair $(\omega,\Phi)$ is regular,
then $\omega\in\cSeq(\Phi)$. If, in addition, $\cSeq(\Phi)=\{\omega\}$, then
$\omega$ is $\varphi$-ergodic since it is  extreme point of the simplex
$\cSI(\fA)$ (see~\cite[Theorem~6.2.44]{Bratteli1981} and its proof). Using the
regularity assumption, one can also argue in the following way. Suppose that
$\omega=\alpha\mu+(1-\alpha)\nu$ for some $\mu,\nu\in\cSI(\fA)$ and
$\alpha\in]0,1[$. The joint convexity of the relative
entropy~\cite[Theorem~1.4]{Ohya1993} yields
\begin{align*}
0=s(\omega|\omega)&\le\alpha s(\mu|\omega)+(1-\alpha)s(\nu|\omega)\\
&=\alpha(-s(\mu)+\mu(E_\Phi)+P(\Phi))+(1-\alpha)(-s(\nu)+\nu(E_\Phi)+P(\Phi))\\
&=-s(\alpha\mu+(1-\alpha)\nu)+(\alpha\mu+(1-\alpha)\nu)(E_\Phi)+P(\Phi)=s(\omega|\omega)=0,
\end{align*}
from which we conclude that $s(\mu|\omega)=s(\nu|\omega)=0$ and hence
$\mu,\nu\in\cSeq(\Phi)=\{\omega\}$.

\noindent 2. For $\nu\in \cSI(\fA)$  and $\omega\in \cSeq(\Phi)$, $\Phi \in
\cBr$, we set
\[
\bar s(\nu|\omega)=\limsup_{\Lambda\uparrow\zz^d}\frac{S(\nu_\Lambda|\omega_\Lambda)}{|\Lambda|}.
\]
By Part~(1) of Proposition~\ref{mil-la} and Relations~\eqref{eq:sDef},
\eqref{eq:def_pressure} and~\eqref{eq:seForm},
\begin{equation}
\bar s(\nu|\omega)\leq-s(\nu) + \nu(E_\Phi) + P(\Phi).
\label{mil-mor}
\end{equation}
Hence, proving regularity reduces to establishing the lower bound 
\[
\liminf_{\Lambda\uparrow\zz^d}\frac{S(\nu_\Lambda|\omega_\Lambda)}{|\Lambda|}\geq 
-s(\nu) + \nu(E_\Phi) + P(\Phi).
\]
Remarkably, Theorem~\ref{thm:weakGibbs} remains the best existing result about this
estimate.

\noindent 3. In~\cite{Hiai2007,Ogata2011}, the authors consider
\textit{asymptotically decoupled} states $\omega\in\cS(\fA)$ defined by the
property
\[
\e^{-c_\Lambda}\omega_\Lambda\otimes\omega_{\Lambda^c}\leq\omega\leq \e^{c_\Lambda}\omega_\Lambda\otimes\omega_{\Lambda^c},
\]
where the $c_\Lambda$ satisfy~\eqref{mil-er}. If in addition
$\omega\in\cSeq(\Phi)$, $\Phi\in\cBr$, the proof of~\cite[Lemma 2.3]{Hiai1993a}
gives that $\omega\in \cSlwg(\Phi)$.

\subsection{Main results} 
\label{sec-main-results}

For $\omega\in\cSI(\fA)$  and  $\Phi\in\cBr$, we set  
\be
\bar \omega_T=\frac{1}{T}\int_0^T \omega \circ \alpha_\Phi^t \d t,
\label{eq:Cesaro}
\ee
and denote by $\cS_+(\omega,\Phi)$ the set of \wstar{limit} points of the net
$\{\bar \omega_T\}_{T>0}$ as $T\rightarrow \infty$. We will call the elements of
$\cS_+(\omega, \Phi)$ \textit{Equilibrium Steady States} (ESS) of
$(\fA,\alpha_\Phi,\omega)$. The set $\cS_+(\omega, \Phi)$ is obviously
non-empty. Since $\fA$ is separable,  $\omega_+\in\cS_+(\omega,\Phi)$ iff there
exists sequence $T_n\uparrow \infty$ such that $\omega_+=\lim_{n\rightarrow
	\infty}\bar\omega_{T_n}$. The invariance properties stated in
Theorem~\ref{ver-tr} survive time-averaging, see
Section~\ref{sec-proof-ver-tr-0}.
\begin{prop}\label{ver-tr-0}
For any $\omega\in\cSI(\fA)$ and any $\Phi\in\cBr$ the following hold for all $T\in\rr$:
\begin{enumerate}[label=(\arabic*)]
	\itemsep0em
\item $s(\omega)=s(\bar \omega_T)$.
\item $\omega (E_\Phi)=\bar \omega_T(E_\Phi)$.
\item In the fermionic case, $\omega(E_N)=\bar\omega_T(E_N)$.
\end{enumerate}
\end{prop}
 
The elements of $\cS_+(\omega, \Phi)$ have the following basic properties:

\begin{prop}\label{ver-tr-2}
Let $\omega_+\in\cS_+(\omega,\Phi)$. Then: 
\begin{enumerate}[label=(\arabic*)]
\itemsep0em
\item $\omega_+\in\cSI(\fA)$ and $\omega_+\circ\alpha_\Phi^t=\omega_+$ for all $t\in\rr$.
\item $s(\omega)\leq s(\omega_+)$.
\item $\omega(E_\Phi)=\omega_+(E_\Phi)$.
\item In the fermionic case, $\omega(E_N)=\omega_+(E_N)$.
\end{enumerate}
\end{prop}
Part~(1) follows from the translation invariance of the time averaged states
$\bar\omega_{T}$ and a well known property on Cesàro averages. Part~(2) follows
from Proposition~\ref{ver-tr-0}-(2) and the upper semi-continuity of the
specific entropy. Parts~(3) and~(4) follows from  Parts~(2) and~(3) of
Proposition~\ref{ver-tr-0}.

\medskip
We now turn to our  main results.
\begin{thm}\label{thm-mil}
Let $\Phi,\Psi\in\cBr$,  $\omega\in \cSeq(\Psi) $, and 
$\omega_+\in\cS_+(\omega,\Phi)$. Then $\omega(E_\Psi)\leq\omega_+(E_\Psi)$.
\end{thm}

The remaining results  depend critically on the regularity assumption.
\begin{thm}\label{thm-main-2}
Let $\Phi,\Psi\in\cBr$. Suppose that $(\omega,\Psi)$ is a regular pair and that
$\omega_+\in\cS_+(\omega,\Phi)$. Then the  following statements are equivalent:
\begin{enumerate}[label=(\alph*)]
\itemsep0em
\item $\omega(E_\Psi)= \omega_+(E_\Psi)$.
\item $\omega_+\in\cSeq(\Psi)$ and $s(\omega)=s(\omega_+)$.
\end{enumerate}
Suppose in addition that $\omega_+\in\cSeq(\Phi)$. Then (a) and (b) are further equivalent to:
\begin{enumerate}[label=(\alph*)]
\itemsep0em
\setcounter{enumi}{2}
\item $\Phi\sim \Psi$.
\item $\omega=\omega_+$.
\end{enumerate}
\end{thm}

\begin{thm}\label{main-thm-1}
Let  $\omega\in\cSI(\fA)$, $\Phi\in\cBr$. For any
$\omega_+\in\cS_+(\omega,\Phi)\cap\cSeq(\Phi)$ one has
$$
s(\omega)=s(\omega_+)\ \Longleftrightarrow\ \omega=\omega_+.
$$ 
\end{thm}

An immediate consequence of these two theorems is:

\begin{cor}\label{main-forg}
Let  $\omega\in\cSI(\fA)$, $\Phi, \Psi \in\cBr$,
$\omega_+\in\cS_+(\omega,\Phi)$, and  suppose that $\omega$ is 
not~$\alpha_\Phi$-invariant.
\begin{enumerate}[label=(\arabic*)]
\itemsep0em
\item If the pair $(\omega, \Psi)$ is regular and $\cSeq(\Psi)=\{\omega\}$, then~$\omega(E_\Psi)<\omega_+(E_\Psi)$.
\item If  $\omega_+\in\cSeq(\Phi)$, then $s(\omega)<s(\omega_+)$.
\end{enumerate}
\end{cor}

The proofs of Theorems~\ref{thm-mil}, \ref{thm-main-2}, and~\ref{main-thm-1} are
given  in Sections~\ref{sec-proof-mil}, \ref{sec-proof-main-2}
and~\ref{sec-proof-main-1}. Although they  are technically not difficult, to the
best of our knowledge they are the first general results  about the strict
increase of specific energy/entropy in quantum statistical mechanics. Central to
the proof of the inequality $\omega(E_\Psi)<\omega_+(E_\Psi)$ is the properly
identified concept of regularity. We emphasize that  the latter property is
quite universal: recall that according to Theorem~\ref{thm:weakGibbs}, every
equilibrium state at high temperature is weak Gibbs and hence is regular. The
inequality $s(\omega)<s(\omega_+)$ is an immediate consequence of energy
conservation $\omega(E_\Phi)=\omega_+(E_\Phi)$ and the Gibbs variational
principle applied to  $\omega_+\in \cSeq(\Phi)$.

\noindent{\bf Remark.} Fixing $\beta=1$ arises differently in the contexts of the
initial state $\omega\in\cSeq(\Psi)$ and the EES $\omega_+\in\cS_+(\omega,\Phi)$. In
the case of $\omega$, $\beta$ is simply absorbed in $\Psi$. On the other hand,
rescaling $\Phi$ as $\gamma\Phi$, $\gamma>0$, does not change EES, and instead
gives a free parameter $\gamma$. Since $\omega_+$ can be in at most one of the
sets $\cSeq(\gamma\Phi)$, $\gamma >0$,\footnote{These sets are disjoint} if it
happens that we have approach to equilibrium and that
$\omega_+\in\cSeq(\bar\gamma \Phi)$ for some $\bar\gamma >0$, rescaling $\Phi$
as $\bar\gamma\Phi$ sets the inverse temperature of $\omega_+$ to $1$. In the
next section we will comment on the choice of sign $\gamma>0$ in this remark.

\subsection{Remarks}
\label{sec-discussion}

\paragraph{Approach to equilibrium in algebraic quantum statistical mechanics.}
As we have already remarked, algebraic quantum statistical mechanics provides a
natural mathematical framework for the study of dynamical aspects of important
classes of infinitely extended quantum systems. In spite of its many successes,
it remains unknown whether this framework, even in principle, can account for
the zeroth law of thermodynamics. To the best of our knowledge, the 
works~\cite{Lanford1968,Hugenholtz1983,Hugenholtz1987}  are the only attempts at
a formulation of the problem. Guided by  more recent works on foundations of
non-equilibrium quantum statistical mechanics in the algebraic framework, we
have presented  here what we believe to be the minimal formulation of the
problem of approach to equilibrium, and have obtained several  structural
results that are physically natural. Much remains to be done in developing a
general structural theory and studying concrete physically relevant models in
the context of this proposal.

We emphasize that Theorem~\ref{main-thm-1} and  the second  parts of
Theorem~\ref{thm-main-2} and  Corollary~\ref{main-forg} are conditional results.
The basic questions when the set $\cS_+(\omega,\Phi)$ of ESS contains $\omega_+$
such that $\omega_+\in\cSeq(\Phi)$ remains open; no non-trivial example is
known.

Theorem~\ref{thm-mil} and the first  part of Theorem~\ref{thm-main-2} and
Corollary~\ref{main-forg} are unconditional results. Their  basic consequence is
the inherent irreversibility of approach to equilibrium. To elucidate this
point, let $\Phi,\Psi\in\cBr$, and suppose that $(\omega,\Psi)$ is a  regular
pair satisfying $\cSeq(\Psi)=\{\omega\}$. Let $\omega_+\in\cS_+(\omega,\Phi)$
and assume reversibility in the sense that $\omega\in\cS_+(\omega_+,\Psi)$.
Then, by Part~(3) of Proposition~\ref{ver-tr-2},
$\omega(E_\Psi)=\omega_+(E_\Psi)$, and  the implication (a) $\Rightarrow$ (b) of
Theorem~\ref{thm-main-2} gives that $\omega_+\in\cSeq(\Psi)=\{\omega\}$. Hence,
$\omega=\omega_+$, and the setting is trivial in the sense that $\omega$ is
$\alpha_\Phi$-invariant. This irreversibility  is in sharp contrast with the
reversibility of return to equilibrium which  we discuss below.

\paragraph{Direction of time.} Changing  the direction of time amounts to
replacing the average~\eqref{eq:Cesaro} with
\[
\bar \omega_T=\frac{1}{T}\int_0^T \omega \circ \alpha_\Phi^{-t} \d t.
\]
We denote by $\cS_-(\omega,\Phi)$ the corresponding set of ESS. Since
$\cS_-(\omega,\Phi)=\cS_+(\omega, -\Phi)$, none of results discussed in this
paper is affected by such a change, as expected; see~\cite[Section 3.8]{Peierls1979}
for a lucid discussion of this point. If the quantum dynamical
system $(\fA, \alpha_\Phi, \omega)$ is time-reversal invariant with
time-reversal $\Theta$\footnote{$\Theta$ is an anti-linear $\ast$-automorphism
of $\fA$ satisfying $\Theta \circ \alpha_\Phi^t=\alpha_\Phi^{-t}\circ \Theta$
for all $t\in \rr$ and $\omega(\Theta(A))=\omega(A^\ast)$ for all $A\in \fA$.}, 
then $\nu\in \cS_-(\omega,\Phi)$ iff $\overline{\nu\circ \Theta} \in \cS_+(\omega,\Phi)$
and $\cS_-(\omega, \Phi)\cap \cSeq(\Phi)=\cS_+(\omega, \Phi)\cap \cSeq(\Phi)$.

\paragraph{Question of Ruelle.}
To the best of our knowledge, the question about the increase of specific
entropy with time was first raised in  1967 by David Ruelle. In~\cite[p.
1666]{Ruelle1967} Ruelle writes:
\begin{quote}\textit{
It is unclear to the author whether the evolution of an infinite system should
increase its entropy per unit volume. Another possibility is that, when the time
tends to $+\infty$, a state has a limit with strictly larger entropy.}
\end{quote}
In the context of quantum spin systems, the first part of the question was
answered a year later by Lanford and Robinson~\cite{Lanford1968}---the specific
entropy remains constant for finite times. One of our main results sheds a light
on the second part of Ruelle's question by establishing the conditional result:
if $\omega_+\in \cSeq( \Phi)$ and  $\omega\not=\omega_+$, then
$s(\omega_+)>s(\omega)$. As discussed above, a parallel to Ruelle's question and
another signature of irreversibility is the inequality $\omega_+(E_\Psi)>
\omega(E_\Psi)$. Our other main result is that this inequality 
unconditionally holds in the high temperature regime and that its general
validity is linked to the concept of regularity.

\paragraph{Comparison with return to equilibrium.}
In spite of a formal similarity, it is important to emphasize that the
properties of return to equilibrium  and approach to equilibrium are very
different. Return to equilibrium is an ergodic property asserting that thermal
equilibrium states are stable under \textit{local perturbations.} Approach to
equilibrium asserts that, under a \textit{global perturbation,} the system
approaches a new equilibrium state. The mechanisms leading to the respective
phenomena are completely different. In return to equilibrium, a local
perturbation $V$ disperses to spatial infinity while the specific entropy
remains constant in the large time limit. It is perfectly possible, and in fact
quite common, that $\omega\not=\omega_V$ and that both dynamical systems
$(\cO,\alpha,\omega)$ and $(\cO,\alpha_V,\omega_V)$ have the property of return
to equilibrium, in which case return to equilibrium is non-trivially reversible
in the sense that
\[
\lim_{T\rightarrow \infty}\frac{1}{T}\int_0^t\omega \circ \alpha_V^t \d t=\omega_V 
\qquad\hbox{and}\qquad 
\lim_{T\rightarrow \infty}\frac{1}{T}\int_0^t\omega_V \circ \alpha^t \d t=\omega.
\]
In comparison, approach to equilibrium is inherently irreversible, as discussed
in the previous remark.

The formal similarity between the return and approach to equilibrium requires a
further comment. Consider a triple $(\fA,\alpha_\Psi,\omega)$, where
$\Psi\in\cBr$ and $\omega$ is a $\alpha_\Psi$-KMS state. Let $\Phi\in\cBr$ and
consider the local perturbation
\[
V_\Lambda=\sum_{X\subset \Lambda} (\Phi(X)- \Psi(X)).
\]
Denoting by $\alpha_\Lambda$ the perturbed $C^\ast$-dynamics, and by
$\omega_{\Lambda}$ the perturbed KMS-state, and assuming that
$(\fA,\alpha_\Lambda,\omega_\Lambda)$ has the property of return to equilibrium,
one has
\[
\lim_{T\rightarrow\infty}\frac{1}{T}\int_0^T\omega\circ\alpha_{\Lambda}^t\d t=\omega_\Lambda.
\]
It is not difficult to show that
$\lim_{\Lambda\uparrow\zz^d}\alpha_\Lambda^t=\alpha_\Phi^t$ strongly, and that
for sufficiently small interactions
$\lim_{\Lambda\uparrow\zz^d}\omega_\Lambda=\omega_+$, where $\omega_+$ is the
unique $\alpha_\Phi$-KMS state. Under these circumstances\footnote{The principal
of which is that $(\fA,\alpha_\Lambda,\omega_\Lambda)$ has the property of
return to equilibrium for all large enough $\Lambda$.} one has
\begin{equation}
\lim_{\Lambda\uparrow\zz^d}\lim_{T\rightarrow\infty}\frac{1}{T}\int_0^T\omega\circ\alpha_\Lambda^t\d t=\omega_+.
\label{fog}
\end{equation}
If the two limits in~\eqref{fog} can be interchanged, then approach to
equilibrium holds. Our main results imply  that this strategy  cannot be reduced
to technical improvements of the existing spectral/scattering estimates used in
the literature to establish return to equilibrium and that novel ideas are
needed.

\paragraph{Comparison with the non-equilibrium theory.} The mathematical theory
of non-equilibrium quantum statistical mechanics developed over the last twenty
years has played an important role in the genesis of this work. In fact, a
strong similarity exists between the conceptual framework of the present paper
and the one developed in~\cite{Jaksic2001a,Ruelle1999}, in the context of a now
mature non-equilibrium quantum statistical program.

The entropy balance equation, Identity~\eqref{ep-ness}, is the central starting
point of the non-equi\-li\-brium theory. The regularity
condition~\eqref{eq:regularity_condition} plays a similar role in the structural
theory of approach to equilibrium developed in this work.

In the construction of Non-Equilibrium Steady States (NESS) proposed in the
above mentioned program, the state of the system at time $t$, given by
$\omega\circ\alpha_V^t$, is normal with respect to the initial state $\omega$.
The NESS, however, are typically singular with respect to $\omega$, and in fact
under suitable regularity assumptions, the strict positivity of entropy
production---a signature of non\-equilibrium--- is equivalent to this
singularity. These regularity assumptions can be either verified  generically or
by a detailed study of concrete models; for discussion of these points we refer
the reader to~\cite{Aschbacher2006, Jaksic2002b, Jaksic2007b}. On the other
hand, in the context of Theorems~\ref{thm-main-2} and~\ref{main-thm-1}, and
recalling the remark after Definition~\ref{def:regular}, if $(\omega,\Psi)$ is a
regular pair such that $\cSeq(\Psi)=\{\omega\}$, then, for any $t$, the state
$\omega\circ \alpha_\Phi^t$ is $\varphi$-ergodic, and  so either  $\omega\circ
\alpha_\Phi^t=\omega$ or the states $\omega\circ \alpha_\Phi^t$ and $\omega$ are
mutually singular. The same holds for the ESS
$\omega_+\in\cS_+(\omega,\Phi)\cap\cSeq(\Phi)$: either $\omega=\omega_+$ or the
states $\omega$ and $\omega_+$ are mutually singular, and in the later case
$s(\omega_+)>s(\omega)$. To summarize, the singularity of the NESS in the
non-equilibrium setting is an essential and  non-trivial dynamical problem. The
singularity of ESS in the approach of equilibrium setting is a direct
consequence of the translation invariance and the general structure of algebraic
quantum statistical mechanics. The generality of our results, in comparison with
related results in non-equilibrium statistical mechanics, can be understood in
these terms.

\paragraph{Weak coupling limit.} The weak coupling (or van Hove) limit sheds
further light on the above remarks. In the context of return to equilibrium,
NESS, and Entropy Production, open quantum systems consisting of small quantum
system with finite dimensional Hilbert space locally coupled to finitely many
free fermionic or bosonic reservoirs---such models  are often called
Pauli--Fierz systems---play a privileged role. The weak coupling limit of  such
models is described by  Pauli's master equation and  the respective theory was
developed  in  seminal works~\cite{Davies1974, Spohn1978b}; for modern
expositions of this theory see~\cite{Derezinski2006, Jaksic2014a}.  The weak
coupling limit theory played a very important  role in more recent studies  of
return of equilibrium. In a similar spirit, the interacting fermion systems
of~\cite{Erdos2004, Hugenholtz1983, Hugenholtz1987} play a privileged role in
study of approach to equilibrium. The weak coupling limit of such models is
formally described  by the quantum Boltzmann equation whose mathematically
rigorous derivation remains an outstanding open problem;
see~\cite{Benedetto2007,Lukkarinen2011,He2021} for a review and important
progresses in this direction.

\paragraph{Weak Gibbsianity and regularity.} In spite of its relative late
appearance, the concept of weak Gibbs state plays an important role in the study
of invariant measures of classical dynamical systems, as a natural boundary for
the validity of the thermodynamic formalism. In the framework of statistical
mechanics it is known that, under very general conditions, the equilibrium
states of classical spin systems with summable interactions are weak
Gibbs\footnote{The converse does not hold.}~\cite{Pfister2020}.
Theorem~\ref{thm:weakGibbs} is a much weaker result. Its proof relies on the 
Gibbs condition (see Proposition~\ref{prop:w-s-1}) and
Theorem~\ref{thm:lenci-luc}, the restrictions coming from the latter result.
Regarding these restrictions, we mention the article~\cite{Moriya2020} where the
proof of regularity of  any pair $(\omega,\Phi)\in\cSeq(\Phi)\times\cBr$ is
announced. Unfortunately, this proof is incomplete and cannot be fixed along the
proposed lines.\footnote{The error concerns the identity~(32)
in~\cite{Moriya2020}, which implies that the KMS vector representative
constructed by Ejima and Ogata in~\cite{Ejima2019} is in the natural cone.
Unfortunately, a simple argument shows that this is so only in the commutative
case.}

A better understanding of the status of weak Gibbs states in equilibrium quantum
statistical mechanics, and more generally quantum spin dynamics, is central to
the success of the proposed research program. We also believe that, similarly to
the classical setting, quantum weak Gibbs states will find many other
applications in quantum statistical mechanics.

\paragraph{$\boldsymbol{W^\ast}$-dynamical systems.} In many important physical
examples, and in particular those involving bosons, it may be more convenient to
work in a $W^\ast$-framework. Similar to the non-equilibrium quantum statistical
mechanics\footnote{See~\cite{Merkli2007,Merkli2007a}.}, it appears difficult to
develop a general structural theory of approach to equilibrium in the
$W^\ast$-setting. In this case one is naturally limited to the study of concrete
models.

 \section{Proofs}
 \label{sec-proofs}

\subsection{Proof of Theorem~\ref{thm:spin-phys-eq}}
\label{sec-phys-eq}

In this section  we prove the parts of Theorem~\ref{thm:spin-phys-eq} that are not in~\cite[Theorem~III.4.2]{Israel1979}.

\medskip\noindent(a) $\Leftrightarrow$ (b). This equivalence\ is a direct consequence of Relation~\eqref{eq:aschball} and the fact that
$$
H_\Lambda(\Phi)-H_\Lambda(\Psi) = H_\Lambda(\Phi-\Psi).
$$

\medskip\noindent(f) $\Leftrightarrow$ (g). By~\cite[Theorem~6.2.4]{Bratteli1981},
if $\Phi\in\cBr$, then $\alpha_\Phi^t=\e^{t\delta}$, where the generating
 derivation $\delta$ is the closure of the map
$$
\fAloc\ni A\mapsto \sum_{Y\cap\supp(A)\neq\emptyset}\i[\Phi(Y),A].
$$
Thus, $\alpha_\Phi=\alpha_\Psi$ iff, for any $A\in\fAloc$,
$$
\sum_{Y\cap\supp(A)\neq\emptyset}\i[\Phi(Y)-\Psi(Y),A]=0.
$$
Since $\i[\Phi(Y)-\Psi(Y),A]=0$ whenever $Y\cap\supp(A)=\emptyset$, 
this condition is equivalent to~(g).

\medskip\noindent(g) $\Rightarrow$ (b). By the simplicity of $\fA$, 
(g) implies that for any $\Lambda$ there exists $C_\Lambda\in\rr$ such that 
$$
\sum_{X\cap\Lambda\neq\emptyset}(\Phi(X)-\Psi(X))=C_\Lambda\one.
$$
Recalling~\eqref{eq:surface-en}, we deduce
\be
\frac1{|\Lambda|}\left(H_\Lambda(\Phi)-H_\Lambda(\Psi)\right)
=\frac{C_\Lambda}{|\Lambda|}\one-\frac1{|\Lambda|}W_\Lambda(\Phi-\Psi),
\label{eq:Alcina}
\ee
while~\eqref{eq:aschball} further yields
$$
\lim_{\Lambda\uparrow\zz^d}\frac1{|\Lambda|}\tr\left(H_\Lambda(\Phi)-H_\Lambda(\Psi)\right)=\tr(E_{\Phi-\Psi}).
$$
Taking the trace on both sides of~\eqref{eq:Alcina} and 
taking~\eqref{eq:NoSurf} into account yields
$$
\lim_{\Lambda\uparrow\zz^d}\frac{C_\Lambda}{|\Lambda|}=\tr(E_{\Phi-\Psi}).
$$
Finally, letting $\Lambda\uparrow\zz^d$ in~\eqref{eq:Alcina} gives~(b). 

\medskip\noindent(b) $\Rightarrow$ (c). For $t\in[0,1]$ 
set $\Phi_t=(1-t)\Psi+t\Phi$ and note that
$$
H_\Lambda(\Phi_t)-H_\Lambda(\Psi)=H_\Lambda(\Phi_t-\Psi)=t\left(H_\Lambda(\Phi)-H_\Lambda(\Psi)\right)
$$
and~(b) imply
$$
\lim_{\Lambda\uparrow\zz^d}\frac1{|\Lambda|}\left(H_\Lambda(\Phi_t)-H_\Lambda(\Psi)\right)=t\tr(E_{\Phi-\Psi})\one.
$$
Thus, given $\delta>0$, and for large enough $\Lambda$,
$$
H_\Lambda(\Psi)+|\Lambda|(t\tr(E_{\Phi-\Psi})-\delta)\one
\le H_\Lambda(\Phi_t)
\le H_\Lambda(\Psi)+|\Lambda|(t\tr(E_{\Phi-\Psi})+\delta)\one.
$$
Invoking the finite-volume Gibbs variational principle~\cite[Proposition~1.10]{Ohya1993}
\be
P_\Lambda(\Phi)=\log\,\tr\e^{-H_\Lambda(\Phi)}
=\max_{\rho\in\cS(\fA_\Lambda)}\left(S(\rho)-\rho\left(H_\Lambda(\Phi)\right)\right),
\label{eq:PLambdaVariational}
\ee
the previous inequalities yield
$$
\left|P_\Lambda(\Phi_t)-P_\Lambda(\Psi)+|\Lambda|t\tr(E_{\Phi-\Psi})\right|\le\delta|\Lambda|.
$$
In view of~\eqref{eq:def_pressure}, dividing both sides of this 
estimate by $|\Lambda|$, taking the thermodynamic limit $\Lambda\uparrow\zz^d$,
and finally letting $\delta\downarrow0$, we deduce
$$
P(\Phi_t)=P(\Psi)-t\tr(E_{\Phi-\Psi}),
$$
which is~(c). 
 
\subsection{Proof of Theorem~\ref{ver-tr}}
\label{sec-proof-invar}

(1) Invariance of the specific entropy goes back
to~\cite[Theorem~5]{Lanford1968}. Even though it may look surprising,  its proof
can be understood as follows. Consider a tiling of $\zz^d$ with a large cube
$\Lambda$, and perturb the interaction $\Phi$ so that all cubes are
disconnected. The entropy of $\omega$ restricted to $\Lambda$ is constant under
the perturbed dynamics because it is a closed finite system by construction.
Finally, the entropy of the perturbed dynamics coincides with the original one
up to a $o(|\Lambda|)$ boundary term whose contribution to the specific entropy
vanishes in the $\Lambda\uparrow\zz^d$-limit.
	
\medskip\noindent(2) We recall again that, by~\cite[Theorem~6.2.4]{Bratteli1981}, $\alpha_\Phi^t=\e^{t\delta}$, where the generator $\delta$ is the closure of the map
$$
\fAloc\ni A\mapsto \sum_{Y\cap\supp(A)\neq\emptyset}\i[\Phi(Y),A].
$$
Since $\omega\circ\alpha_\Phi^t\in\cSI(\fA)$, it suffices to show that $E_\Phi\in\Dom(\delta)$ and $\omega(\delta(E_\Phi))=0$ for all $\omega\in\cSI(\fA)$.
	
For latter reference, we will show that $E_\Psi\in\Dom(\delta)$ holds for all $\Psi\in\cBr$. Setting
$$
E_{\Psi,n}=\sum_{X\ni0\atop\diam(X)\le n}\frac{\Psi(X)}{|X|}\in\fAloc,
$$
and observing that
$$
\sum_{n=0}^\infty \sum_{X\ni0\atop\diam(X)=n}\frac{\|\Psi(X)\|}{|X|}\le\|\Psi\|_r<\infty,
$$
we have
$$
\lim_{n\to\infty}\|E_{\Psi}-E_{\Psi,n}\|\le\lim_{n\to\infty}\sum_{X\ni0\atop\diam(X)>n}\frac{\|\Psi(X)\|}{|X|}=0.
$$
Moreover, for $m\le n$, the formula
$$
\delta(E_{\Psi,n})-\delta(E_{\Psi,m})
=\sum_{X\ni0\atop m<\diam(X)\le n}\frac1{|X|}\sum_{Y\cap X\neq\emptyset}\i[\Phi(Y),\Psi(X)],
$$
gives the estimate
\begin{align*}
\|\delta(E_{\Psi,n})-\delta(E_{\Psi,m})\|
&\le2\sum_{X\ni0\atop m<\diam(X)\le n}\frac{\|\Psi(X)\|}{|X|}\sum_{Y\cap X\neq\emptyset}\|\Phi(Y)\|\\
&\le2\sum_{X\ni0\atop\diam(X)>m}\|\Psi(X)\|\frac{1}{|X|}\sum_{x\in X}\sum_{Y\ni x}\|\Phi(Y)\|
\le2\|\Phi\|_r\sum_{X\ni0\atop\diam(X)>m}\|\Psi(X)\|,
\end{align*}
which shows that the sequence $(\delta(E_{\Psi,n}))_{n\in\nn}$ is Cauchy. It follows that  $E_\Psi\in\Dom(\delta)$ and
\be
\delta(E_\Psi)=\sum_{X\ni0}\frac1{|X|}\sum_{Y\cap X\neq\emptyset}\i[\Phi(Y),\Psi(X)].
\label{eq:deltaEPhi}
\ee
	
To proceed, we equip $X\in\cF$ with the lexicographic order observing that translation invariance implies $\min(X+x)=\min(X)+x$ for any $X\in\cF$ and $x\in\zz^d$. Set
$$
\cF_0=\{X\in\cF\mid\min(X)=0\},
$$
and note that the set of translates of $X\in\cF_0$ containing $0$ is
$$
\cF_X=\{X-x\mid x\in X\}.
$$
Finally, for $X,Y\in\cF_0$, let
$$
Z(X,Y)=\{z\in\zz^d\mid X\cap(Y+z)\neq\emptyset\}.
$$
	
Relation~\eqref{eq:deltaEPhi} thus writes
$$
\delta(E_\Phi)=\sum_{X\in\cF_0}\frac1{|X|}\sum_{x\in X}
\sum_{Y\cap(X-x)\neq\emptyset}\i[\Phi(Y),\varphi^{-x}(\Phi(X))].
$$
Now for $Y\in\cF$ one has $Y=Y_0+y$, with $Y_0\in\cF_0$ and $y=\min(Y)$. Moreover,
$$
\emptyset\neq Y\cap(X-x)=(Y_0+y)\cap(X-x)=((Y_0+x+y)\cap X)-x
$$
iff $x+y\in Z(X,Y_0)$, so that
$$
\delta(E_\Phi)=\sum_{X\in\cF_0}\frac1{|X|}\sum_{x\in X}
\sum_{Y\in\cF_0}\sum_{x+y\in Z(X,Y)}
\i[\varphi^{y}(\Phi(Y)),\varphi^{-x}(\Phi(X))].
$$
It follows that, for $\omega\in\cSI(\fA)$,
\begin{align*}
\omega(\delta(E_\Phi))
&=\sum_{X\in\cF_0}\frac1{|X|}\sum_{x\in X}
\sum_{Y\in\cF_0}\sum_{z\in Z(X,Y)}
\omega(\i[\varphi^{z}(\Phi(Y)),\Phi(X)])\\
&=\sum_{X,Y\in\cF_0}\sum_{z\in Z(X,Y)}
\omega(\i[\varphi^{z}(\Phi(Y)),\Phi(X)]).
\end{align*}
Symmetrizing the inner sum and observing that $Z(Y,X)=-Z(X,Y)$ yields
\begin{align*}
\omega(\delta(E_\Phi))
&=\frac12\sum_{X,Y\in\cF_0}\sum_{z\in Z(X,Y)}
\omega(\i[\varphi^{z}(\Phi(Y)),\Phi(X)]+\i[\varphi^{-z}(\Phi(X)),\Phi(Y)])\\
&=\frac12\sum_{X,Y\in\cF_0}\sum_{z\in Z(X,Y)}
\omega(\i[\varphi^{z}(\Phi(Y)),\Phi(X)]-\i[\varphi^{z}(\Phi(Y)),\Phi(X)])=0.
\end{align*}

\medskip\noindent(3) In a similar spirit, it suffices to show that $E_N\in\Dom(\delta)$, with $\omega(\delta(E_N))=0$ for all $\omega\in\cSI(\fA)$. The generator $\gamma$ of the gauge group is the closure of the map
$$
\fAloc\ni A\mapsto\sum_{x\in\supp(A)}\i[a_x^\ast a_x,A]
=\sum_{x\in\supp(A)}\i[\varphi^x(E_N),A].
$$
Since any state $\omega\in\cSI(\fA)$ is gauge-invariant\footnote{Any state $\omega\in\cSI(\AAA)$ restricts to the gauge-invariant state $\bar\omega=\int_0^{2\pi}\omega\circ\vartheta^\theta\frac{\d\theta}{2\pi}\in\cSI(\fA)$.}, one has $\omega(\gamma(A))=0$ for any $A\in\Dom(\gamma)$, and in particular
$$
0=\omega(\gamma(\Phi(X)))=\sum_{x\in X}\omega(\i[\varphi^x(E_N),\Phi(X)])
=\sum_{x\in X}\omega(\i[E_N,\varphi^{-x}(\Phi(X))])
=-\sum_{Y\in\cF_X}\omega(\i[\Phi(Y),E_N]).
$$
for all $X\in\cF$. We conclude that
$$
\omega(\delta(E_N))=\sum_{X\ni0}\omega(\i[\Phi(X),E_N])
=\sum_{X\in\cF_0}\sum_{Y\in\cF_X}\omega(\i[\Phi(Y),E_N])=0.
$$
\subsection{Proof of Proposition~\ref{prop:weakGibbs_are_eq}}
\label{sec:proofweakGibbs_are_eq}

\medskip\noindent(1) Is merely a reformulation of the defining relations~\eqref{eq:lwgDef} and~\eqref{mil-er} which will be convenient in the proofs of the remaining statements.

\medskip\noindent(2) By the operator monotonicity of the logarithm~\cite[Example~11.16]{Petz2008}, the inequalities~\eqref{eq:lwgDef} follow from~\eqref{eq:ewgDef}, which proves the first inclusion. To deal with the second one, we note that
$$
\log\omega_{\Phi,\Lambda}^c=-H_\Lambda(\Phi)-P_\Lambda(\Phi),
$$
so that any $\omega\in\cSlwg(\Phi)$ satisfies
$$
- c_\Lambda\le\log\omega_\Lambda+H_\Lambda(\Phi)+P_\Lambda(\Phi)\le c_\Lambda,
$$
where $c_\Lambda$ is such that~\eqref{mil-er} holds. Multiplying with $\omega_\Lambda$ and taking the trace yields 
$$
|S(\omega_\Lambda)-\omega(H_\Lambda(\Phi))-P_\Lambda(\Phi)|\le c_\Lambda.
$$
Dividing by $|\Lambda|$, letting $\Lambda\uparrow\zz^d$ and invoking~\eqref{eq:sDef}, \eqref{eq:def_pressure} and~\eqref{eq:seForm} gives
$$
P(\Phi)=s(\omega)-\omega(E_\Phi),
$$
and hence $\omega\in\cSeq(\Phi)$.

\medskip\noindent(3) If $\omega\in\cSlwg(\Phi)\cap\cSlwg(\Psi)$, then
$$
\limsup_{\Lambda\uparrow\zz^d}|\Lambda|^{-1}\|\log\omega_\Lambda-\log\omega_{\Phi,\Lambda}^c\|=
\limsup_{\Lambda\uparrow\zz^d}|\Lambda|^{-1}\|\log\omega_\Lambda-\log\omega_{\Psi,\Lambda}^c\|=0.
$$
It follows that
$$
\limsup_{\Lambda\uparrow\zz^d}|\Lambda|^{-1}\|\log\omega^\mathrm{c}_{\Phi,\Lambda}-\log\omega^\mathrm{c}_{\Psi,\Lambda}\|=0,
$$
which, in view of~(1), immediately leads to $\cSlwg(\Phi)=\cSlwg(\Psi)$.

\medskip\noindent(4) Let $\omega\in\cSlwg(\Phi)$, and suppose that $\Psi\sim\Phi$. By definition one has
\be
\limsup_{\Lambda\uparrow\zz^d}|\Lambda|^{-1}\|\log\omega_\Lambda-\log\omega_{\Phi,\Lambda}^\mathrm{c}\|=0.
\label{eq:Cavaradossi}
\ee
Theorem~\ref{thm:spin-phys-eq}-(b) further implies that, for any $\delta>0$,
\be
|\Lambda|(\tr(E_{\Phi-\Psi})-\delta)\one\le
H_\Lambda(\Phi)-H_\Lambda(\Psi)\le|\Lambda|(\tr(E_{\Phi-\Psi})+\delta)\one
\label{eq:Tosca}
\ee
provided $\Lambda$ is large enough. Writing the variational formula~\eqref{eq:PLambdaVariational} as
$$
P_\Lambda(\Psi)=\max_{\rho\in\cS(\fA_\Lambda)}\left(S(\rho)-\rho\left(H_\Lambda(\Phi)\right)
+\rho\left(H_\Lambda(\Phi)-H_\Lambda(\Psi)\right)\right),
$$
the inequalities~\eqref{eq:Tosca} allow us to deduce
$$
|\Lambda|(-\tr(E_{\Phi-\Psi})-\delta)\le P_\Lambda(\Phi)-P_\Lambda(\Psi)\le|\Lambda|(-\tr(E_{\Phi-\Psi})+\delta).
$$
Adding the last inequalities to~\eqref{eq:Tosca} gives
$$
\|\log\omega_{\Phi,\Lambda}^\mathrm{c}-\log\omega_{\Psi,\Lambda}^\mathrm{c}\|\le 2\delta|\Lambda|.
$$
Combining this estimate with~\eqref{eq:Cavaradossi} yields
$$
\limsup_{\Lambda\uparrow\zz^d}|\Lambda|^{-1}\|\log\omega_\Lambda-\log\omega_{\Psi,\Lambda}^\mathrm{c}\|\le2\delta,
$$
and taking $\delta\downarrow0$ allows us to conclude that $\omega\in\cSlwg(\Psi)$.

Reciprocally, if $\omega\in\cSlwg(\Phi)=\cSlwg(\Psi)$, then it follows from~(1) that $\omega\in\cSeq(\Phi)\cap\cSeq(\Psi)$
and Theorem~\ref{thm:spin-phys-eq}-(d) yields $\Phi\sim\Psi$.

\subsection{Proof of Theorem~\ref{thm:weakGibbs}}
\label{secAraki}
We will make use of~\cite[Theorem 3.7]{Lenci2005}, which we state and prove in the general setting of Section~\ref{sec-setting}.
 
\begin{thm}\label{thm:lenci-luc}
Let $\omega$ be a faithful $(\alpha,1)$-KMS state and $\omega_V$ the perturbed $(\alpha_V,1)$-KMS state induced by $V=V^\ast\in\cO$.
\begin{enumerate}[label=(\roman*)]
\item If the map
\begin{equation}
\rr \ni t \mapsto \alpha^t(V)\in\cO
\label{eq:lr-map}
\end{equation}
has an analytic extension to the strip $0<\Im z<1/2$ which is bounded and continuous on its closure, then 
\begin{equation*} 
\omega_V\leq C_V\omega,
\end{equation*}
where $C_V=\e^{\|V\| + \|\alpha^{\i/2}(V)\|}$.
\item If the map~\eqref{eq:lr-map} has an analytic extension to the strip  $|\Im z|<1/2$ which is bounded and continuous on its closure, then we also have a lower bound 
$$
\omega_V\geq  D_V\omega,
$$
where $D_V=\e^{-\|V\|- \|\alpha_{-V}^{\i/2}(V)\|}$.
\end{enumerate}
\end{thm}

\noindent{\bf Proof.} For later reference, we give here a proof that is different from the original argument in~\cite{Lenci2005} and emphasizes the role of the modular structure. We will freely use the notation and results of  modular theory and perturbation of the KMS-structure discussed  in~\cite{Derezinski2003a}. In particular, $\mathrm{E}_V^\alpha$ denotes the Araki--Dyson expansional associated to $V$,
\be
\mathrm{E}_V^\alpha(t)=\sum_{n\ge0}(\i t)^n
\int\limits_{0\le s_n\le\cdots\le s_1\le 1}\alpha^{ts_n}(V)\cdots\alpha^{ts_1}(V)\d s_1\cdots\d s_n,
\label{eq:vespro}
\ee
see~\cite[Section 3.1]{Derezinski2003a}.

\medskip\noindent(i) Passing to the GNS representation $(\cH, \pi, \Omega)$ induced by $\omega$, we need to prove that for any $A>0$ in $\cO$,
\[
\omega_V(A)=\frac{\langle\Omega_V,\pi(A)\Omega_V\rangle}{\|\Omega_V\|^2}\leq C_V\langle\Omega,\pi(A)\Omega\rangle,
\]
where $\Omega_V=\pi(\mathrm{E}_V^\alpha(\i/2))\Omega$.
Since $\|\Omega_V\|\geq\e^{-\|V\|/2}$ by the Peierls--Bogoliubov inequality, it suffices to prove that 
\begin{equation} 
\sup_{\cO\ni A>0}\frac{\langle \Omega_V, \pi(A)\Omega_V\rangle}{\langle \Omega, \pi(A)\Omega\rangle }
\leq \e^{\|\alpha^{\i/2}(V)\|}.
\label{eq:so-t}
\end{equation}
From the fact that $J\Omega_V=\Omega_V$ we get
\[
\langle\Omega_V,\pi(A)\Omega_V\rangle=\|\pi(A^{1/2})J\pi(\mathrm{E}_V^\alpha(\i/2))\Omega\|^2
=\|\pi({\rm E}_V^\alpha(\i/2))J\pi(A^{1/2})\Omega\|^2,
\]
and $\langle\Omega,\pi(A)\Omega\rangle=\|J\pi(A^{1/2})\Omega\|^2$ yields 
\[
\sup_{\cO\ni A>0}\frac{\langle\Omega_V,\pi(A)\Omega_V\rangle}{\langle\Omega,\pi(A)\Omega\rangle}
=\|\pi({\rm E}_V^\alpha(\i/2))\|^2,
\]
so that~\eqref{eq:so-t} easily follows from~\eqref{eq:vespro}. 

\medskip\noindent(ii) We infer from the relation $\alpha_{-V}^t(V)=\mathrm{E}_{-V}^\alpha(t)\alpha^t(V)\mathrm{E}_{-V}^\alpha(t)^\ast$ that $\alpha_{-V}^z(V)= \mathrm{E}_{-V}^\alpha(z)\alpha^z(V)\mathrm{E}_{-V}^\alpha(\bar z)^\ast$
is analytic in the strip $0<\Im z<1/2$ and bounded and continuous on its closure. Part~(i) and the relation $\omega=(\omega_{-V})_{V}$ yield the claim.
\hfill$\qed$

\medskip
We now proceed with the proof of Theorem~\ref{thm:weakGibbs}. To simplify the notation, we write $\alpha$ and $W_\Lambda$  for $\alpha_\Phi$ and $W_\Lambda(\Phi)$.

We only need to prove  that $\cSeq(\Phi)\subset\cSewg(\Phi)$. To this end, let $\omega\in\cSeq(\Phi)$. By Theorem~\ref{thm:lenci-luc}, if the surface energy $W_\Lambda$ is $\alpha$-analytic in the strip $|\Im z|<a$ for some $a>1/2$, then 
$$
\e^{-\|W_\Lambda\|-\|\alpha^{\i/2}(W_\Lambda)\|}\le\frac{\omega(A)}{\omega_{-W_\Lambda}(A)}
\le\e^{\|W_\Lambda\|+\|\alpha_{-W_\Lambda}^{\i/2}(W_\Lambda)\|}
$$
holds for any $A\in\fA_\Lambda$ such that $A>0$. By~\eqref{eq:NoSurf} and Proposition~\ref{prop:w-s-1}, to prove that $\omega\in\cSewg(\Phi)$ it is then sufficient to show that
\begin{equation}
\lim_{\Lambda\uparrow\zz^d}\frac1{|\Lambda|}\|\alpha^{\i/2}(W_\Lambda)\|=0,\qquad 
\lim_{\Lambda\uparrow\zz^d}\frac1{|\Lambda|}\|\alpha_{-W_\Lambda}^{\i/2}(W_\Lambda)\|=0.
\label{eq:marseille-st}
\end{equation}
In the case~(a), this follows from the celebrated estimate of
Araki~\cite{Araki1969}, see also~\cite[Proposition 3.9]{Lenci2005} and
discussion after it, and \cite[Proposition 2.4]{Matsui2003a}. In the case~(b) we
argue as follows. Let
\[
a=\frac{r}{2 \|\Phi\|_r} >1/2.
\]
Then, by Theorem~\ref{thm:y-k-1}, for any $X\in\cF$ the map $\rr\ni t\mapsto\alpha^t(\Phi(X))$ has analytic extension to the strip $|\Im z|<a$ such that
\be
\|\alpha^{\i/2}(\Phi(X))\|\leq\|\Phi(X)\|\frac{\e^{r|X|}}{1-\|\Phi\|_r/r}.
\label{eq:Dorilla}
\ee
This estimate gives that the map $\rr\ni t\mapsto\alpha^t(W_\Lambda)$ also has an analytic extension to the strip $|\Im z|<a$ and that 
\[
\|\alpha^{\i/2}(W_\Lambda)\|\leq\sum_{X\cap\Lambda\not=\emptyset\atop X\cap\Lambda^c\not=\emptyset}
\|\Phi(X)\|\frac{\e^{r|X|}}{1-\|\Phi\|_r/r}.
\]
Thus, to establish the first limit in~\eqref{eq:marseille-st} it suffices to show that 
$$
\lim_{\Lambda\uparrow\zz^d}\frac1{|\Lambda|}\sum_{X\cap\Lambda \not=\emptyset\atop X\cap\Lambda^c\not=\emptyset}
\|\Phi(X)\|\e^{r| X|}=0.
$$
This relation is immediate for $\Phi\in\cBf$. The general case follows from the density of $\cBf$ in $\cBr$ and the bound
\[
\frac1{|\Lambda|}\sum_{X\cap\Lambda\not=\emptyset}\|\Psi(X)\|\e^{r|X|}\leq\e^r\|\Psi\|_r
\]
that holds for all $\Psi\in\cBr$. 

To deal with the second limit in~\eqref{eq:marseille-st} we observe that the perturbed dynamics $\alpha_{-W_\Lambda}$ is associated with the non-translation-invariant interaction $\Phi_\Lambda$ given by
$$
\Phi_\Lambda(X)=\begin{cases}
0&\text{if }X\cap\Lambda\neq\emptyset\text{ and }X\cap\Lambda^c\neq\emptyset;\\
\Phi(X)&\text{otherwise}.
\end{cases}
$$
By the Remark after Theorem~\ref{thm:y-k-1}, and in view of the obvious fact that $\|\Phi_\Lambda\|_r\leq \|\Phi\|_r$, the estimate~\eqref{eq:Dorilla} holds with $\alpha_{-W_\Lambda}$ replacing $\alpha$. We can then argue as before. 

\subsection{Proof of Proposition~\ref{prop:w-s-1}}
\label{sec-proof-wg}

Theorem~\ref{thm:spin-Gibbs-con} gives
\begin{equation}
\omega_{-W_\Lambda(\Phi)}=\omega_{\Phi,\Lambda}^\mathrm{c}.
\label{ajde-wg}
\end{equation}
If $\omega\in\cSewg(\Phi)$, then for any $A\in \fA_\Lambda$ satisfying  $A>0$ one has
$$
\e^{-c_\Lambda}\leq\frac{\omega(A)}{\omega_{-W_\Lambda}(A)}\leq \e^{c_\Lambda},
$$
and~\eqref{ajde} follows.

To prove the converse statement, for $A\in\fA_\Lambda$ satisfying $A>0$, the definitions of $\underline{d}_\Lambda$ and $\overline{d}_\Lambda$ and Relation~\eqref{ajde-wg} give 
$$
\underline{d}_\Lambda\omega_{\Phi,\Lambda}^\mathrm{c}(A)
\leq\omega(A) 
\leq \overline{d}_\Lambda\omega_{\Phi,\Lambda}^\mathrm{c}(A).
$$
Setting $c_\Lambda=\log \max(\overline{d}_\Lambda, 1/\underline{d}_\Lambda)$, 
one deduces from~\eqref{ajde} that $\omega\in\cSewg(\Phi)$.

\subsection{Proof of Proposition~\ref{ver-tr-0}}
\label{sec-proof-ver-tr-0}

Parts~(2) and~(3)  are immediate consequences of the corresponding statements of Theorem~\ref{ver-tr}.  

To prove Part~(1), writing the integral~\eqref{eq:Cesaro} as a \wstar{limit} of Riemann sums and using that the entropy map is affine and upper-semicontinuous, we derive the inequality
$$
s(\bar\omega_T)\ge\limsup_{N\to\infty}\sum_{k=0}^{N-1}\frac1N s(\omega\circ\alpha_\Phi^{kt/N}).
$$
Part~(1) of Theorem~\ref{ver-tr} further gives $s(\omega\circ\alpha_\Phi^{kt/N})=s(\omega)$, so that $s(\bar\omega_T)\geq s(\omega)$. To prove the reverse inequality, note first that the identity, 
\[
\bar\omega_T=\frac12\bar\omega_{T/2}+\frac12\bar\omega_{T/2}\circ\alpha_\Phi^{T/2},
\]
the affine property of the specific entropy, and Part~(1) of Theorem~\ref{ver-tr} give that $s(\bar\omega_{T})=s(\bar\omega_{T/2})$. Consequently, $s(\bar\omega_{T})=s(\bar\omega_{T/2^n})$ for any $n\in\nn$, and invoking again the upper-semicontinuity, we conclude
\[
s(\bar \omega_T)=\lim_{n\rightarrow \infty}s(\bar \omega_{T/2^n})\leq s(\omega).
\]

\subsection{Proof of Theorem~\ref{thm-mil}}
\label{sec-proof-mil}
By Part~(1) of Proposition~\ref{ver-tr-0}, $s(\bar \omega_T)=s(\omega)$. Relation~\eqref{mil-mor} and the non-negativity of relative entropy give 
\begin{align}
0\leq \bar s(\bar\omega_T|\omega)&\leq -s(\bar\omega_T)+\bar\omega_T(E_\Psi)+P(\Psi)\nonumber\\
&=-s(\omega)+\bar\omega_T(E_\Psi)+P(\Psi)\label{eq:start-mil}\\
&=\bar\omega_T(E_\Psi)-\omega(E_\Psi).\nonumber
\end{align}
Pick a sequence $(T_n)$ such that $\bar\omega_{T_n}\rightarrow\omega_+$. Taking the limit along this sequence in~\eqref{eq:start-mil}, we derive that $\omega_+(E_\Psi)\geq\omega(E_\Psi)$. 
\subsection{Proof of Theorem~\ref{thm-main-2}}
\label{sec-proof-main-2}
Since our hypothesis implies that $\omega\in\cSeq(\Psi)$, we can write
\begin{align}
0\leq s(\bar\omega_T|\omega)&= -s(\bar\omega_T)+\bar\omega_T(E_\Psi)+P(\Psi)\nonumber\\
&=-s(\omega)+\bar\omega_T(E_\Psi)+P(\Psi)\label{eq:start}\\
&=\bar\omega_T(E_\Psi)-\omega(E_\Psi).\nonumber
\end{align}
Pick again a sequence $(T_n)$ such that $\bar\omega_{T_n}\rightarrow\omega_+$ and note that  Relation~\eqref{eq:regularity_condition} gives that the map $\cSI(\fA)\ni\nu\mapsto s(\nu|\omega)$ is lower-semicontinuous. Hence, Relation~\eqref{eq:start} further yields
\[
0\le s(\omega_+|\omega)
\le\liminf_{n\to\infty}s(\bar\omega_{T_n}|\omega)
=\omega_+(E_\Psi)-\omega(E_\Psi).
\]
Thus, if $\omega_+(E_\Psi)=\omega(E_\Psi)$, then $s(\omega_+|\omega)=0$ and, in view of the fact that $\omega_+\in\cSI(\fA)$, \eqref{eq:regularity_condition} gives
\be
0=s(\omega_+|\omega)=-s(\omega_+)+ \omega_+(E_\Psi) + P(\Psi).
\label{eq:glas}
\ee
This allows us to conclude that $\omega_+\in\cSeq(\Psi)$. Since $\omega\in\cSeq(\Psi)$, we also have 
$$
0=-s(\omega)+\omega(E_\Psi)+P(\Psi)=-s(\omega)+\omega_+(E_\Psi)+P(\Psi)
$$
which, upon comparison with~\eqref{eq:glas}, shows that $s(\omega)=s(\omega_+)$, and so (a)~$\Rightarrow$~(b). The implication (b)~$\Rightarrow$~(a) is obvious.

If in addition $\omega_+\in\cSeq(\Phi)$, then (b) gives that $\omega_+\in\cSeq(\Phi)\cap\cSeq(\Psi)$ and~(c) 
follows from  Theorem~\ref{thm:spin-phys-eq}-(d).

Finally, if $\Phi\sim \Psi$, then $\alpha_\Phi=\alpha_\Psi$ by Theorem~\ref{thm:spin-phys-eq}-(f), and so $\omega$ is $\alpha_\Phi$-invariant. Hence, $\omega=\omega_+$, and so (c)~$\Rightarrow$~(d). Obviously (d)~$\Rightarrow$~(a).

\subsection{Proof of Theorem~\ref{main-thm-1}}
\label{sec-proof-main-1}
Suppose that $s(\omega)=s(\omega_+)$. Then, since $\omega_+(E_\Phi)=\omega(E_\Phi)$, 
\begin{align*} 
0&=-s(\omega_+)+\omega_+(E_\Phi)+P(\Phi)\\
&=-s(\omega)  +\omega(E_\Phi)+P(\Phi),
\end{align*}
and so $\omega\in \cSeq(\Phi)$. This gives that  $\omega$ is $\alpha_\Phi$-invariant and that  $\omega=\omega_+$.

\bibliographystyle{capalpha}
\bibliography{MASTER}

\end{document}